\documentclass[sigconf]{acmart}
\acmConference[ESEC/FSE 2018]{Submitted to 11th Joint Meeting of the European Software Engineering Conference and the ACM SIGSOFT Symposium on the Foundations of Software Engineering}{4--9 November, 2017}{Florida, United States}
 
\usepackage{wrapfig}
\usepackage{booktabs} 
\usepackage{algorithmicx} 
\usepackage[linesnumbered,ruled,vlined]{algorithm2e}
\SetKwInOut{Parameter}{Parameter}
\SetKwInOut{Require}{Require Func}
\SetKwInOut{Input}{Input}
\SetKwInOut{Output}{Output}
\usepackage{algpseudocode}
\usepackage{multirow}
\usepackage{tabu}
\usepackage{enumitem} 
\usepackage{tabularx}

\newcolumntype{s}{>{\centering \arraybackslash \hsize=.5\hsize}X}  

\usepackage[framemethod=tikz]{mdframed}
\usepackage{lipsum}
\usetikzlibrary{shadows}
\usepackage{mathtools}
\usepackage{balance}
\usepackage{graphics}
\newmdenv[
tikzsetting= {fill=white},
linewidth=1pt,
roundcorner=2pt, 
shadow=false
]{myshadowbox}

\usepackage{color,colortbl}
\usepackage{xcolor}
\definecolor{lightgray}{gray}{0.8}

\makeatletter
\let\th@plain\relax
\makeatother

\usepackage[framed]{ntheorem}
\usepackage{framed}
\usepackage{tikz}
\usetikzlibrary{shadows}

\definecolor{Gray}{rgb}{0.88,1,1}
\definecolor{Gray}{gray}{0.85}
\theoremclass{Lesson}
\theoremstyle{break}
\tikzstyle{thmbox} = [rectangle, rounded corners, draw=black,
 fill=Gray,  drop shadow={fill=black, opacity=1}]

\newcommand{\wei}[1]{\textcolor{red}{TODO: #1}}

\newcommand{\bi}{\begin{itemize}[leftmargin=0.4cm]}
\newcommand{\ei}{\end{itemize}}
\newcommand{\be}{\begin{enumerate}}
\newcommand{\ee}{\end{enumerate}}
\newcommand{\tion}[1]{\S\ref{sect:#1}}
\newcommand{\fig}[1]{Figure~\ref{fig:#1}}

\newshadedtheorem{lesson}{Result}

\setcopyright{rightsretained}

\setcopyright{acmcopyright}
\acmDOI{10.1145/3106237.3106256}
\acmISBN{978-1-4503-5105-8/17/09}
\acmConference[ESEC/FSE'18]{2018 11th Joint Meeting of the European Software Engineering Conference and the ACM SIGSOFT Symposium on the Foundations of Software Engineering}{November 4--9, 2018}{Florida, United States} 
\acmYear{2018}
\copyrightyear{2018}
\acmPrice{15.00}
\begin{document} 
\title{Building Better Quality Predictors Using ``$\epsilon$-Dominance'' }

\author{Wei Fu, Tim Menzies, Di Chen and Amritanshu Agrawal}
\affiliation{%
  \institution{ Com. Sci., NC State, USA}
}
\email{{wfu, dchen20, aagrawa8}@ncsu.edu,tim.menzies@gmail.com}

 
\begin{abstract}
Despite extensive research, many methods in software quality prediction still exhibit some degree
of  uncertainty in their results.
Rather than treating this  as a problem, this paper asks if this uncertainty is
a  resource that can 
simplify software quality prediction. 

For example, 
Deb's principle of $\epsilon$-dominance
  states that if there exists some
$\epsilon$
value below which
it is useless or impossible to distinguish results, then
it is superfluous to explore anything less than
$\epsilon$. We
say that for ``large $\epsilon$ problems'', the
 results space of learning effectively contains just a few regions.
If many learners are then applied to such large $\epsilon$ problems,  
they would exhibit a ``many roads lead to Rome'' property; i.e.,  many different software quality prediction methods
  would generate a small set of very similar results.

This paper explores DART,  an algorithm especially selected to succeed for
large $\epsilon$ software quality prediction problems. DART is remarkable simple
yet, on experimentation, it
dramatically out-performs three sets of state-of-the-art defect prediction
methods. 

The success of DART  for defect prediction
begs the questions: how many other domains in software quality predictors
can also be radically simplified? This will be a fruitful direction
for future work.

\end{abstract}


\maketitle


\section{Introduction}

This paper presents DART,  a  novel method
for simplifying  supervised
learning for defect prediction.  DART produces tiny, easily comprehensible models (5 lines of very simple rules)
 and, 
 in principle, DART could be applied to many domains in software quality predictors.
 When tested on software defect prediction, this  method dramatically out-performs
 three recent state-of-the-art studies~\cite{ghotra2015revisiting,fu2016tuning, agrawal2018better}.

DART was designed by working backward from known properties
of   
software quality predictors problems.
Such predictors  exhibit a ``many roads  lead to Rome'' property;
i.e.,  many different data mining
algorithms generate a small set of very similar results.
For example,   Lessmann et al. reported that 17 of 22 studied data mining algorithms for defect prediction had statistically indistinguishable performance ~\cite{Lessmann08}.
Also,
Ghotra et al. reported that the performance of 32 data mining algorithms for defect prediction clustered into just four groups~\cite{ghotra2015revisiting}. 

This paper asks what can be learned from the above examples.
In this paper, we  note that  learners that have a ``results space''
i.e., values for various performance metrics such as recall and false alarm.
Next, we ask what ``shape'' of result spaces leads to  ``many roads''? Also, given those ``shapes'', do we need complex data miners? Or can we reverse engineer from that space a much simpler
kind of software quality predictor?

To answer these questions we apply  $\epsilon$-dominance~\cite{deb2005evaluating}.
Deb's principle of $\epsilon$-dominance
  states that if there exists some
$\epsilon$
value below which
it is useless or impossible to distinguish results, then
\begin{quote}
 {\bf {\em It is superfluous to explore anything less than $\epsilon$.}}
\end{quote}
We say that for ``large $\epsilon$ problems'', the results space of  learning
effectively contains just a few regions
In such simple result spaces, a few  DARTs thrown around the output space
would sample the results  just as well, or better, than more complex methods.

To test if $\epsilon$-dominance simplifies
software quality prediction,
this paper compares DART-ing around the results
space 
against three defect prediction systems:
\begin{enumerate}
\item
The   algorithms surveyed at a recent  ICSE'15 paper~\cite{ghotra2015revisiting};
\item
A hyper-parameter optimization method  proposed in 2016 in the IST journal~\cite{fu2016tuning}; 
\item
A search-based data pre-processing method presented at ICSE'18~\cite{agrawal2018better}. 
\end{enumerate}
These three were chosen since they reflect the state-of-the-art in software quality defect prediction.
Also, the second and third items in this list are  CPU-intensive systems that require days of computing time to execute
data algorithms many times to find good configurations. 
Comparing something as simple
as DART to these
complex systems let us critically assess the value of elaborate cloud computing environments for software
quality prediction.

What we will see is that  a small number of DARTs 
 dramatically out-performs these three systems.
This suggests that, at least for our data,  much of the complexity
 associated with  hyper-parameter optimization is not required.
We conjecture that
 a few  DARTs succeed so well since the results space for defect prediction exhibits the large $\epsilon$ property.
We also conjecture that
prior state-of-the-art algorithms fail against DART since all those models
do not spread out over the results space. On the other hand, DART works so well since it knows how to spread its
models across a large $\epsilon$ results space better.

\begin{table*}[!hpt]
\renewcommand{\baselinestretch}{1}
\scriptsize
\caption{OO Measures used in our defect data sets.}\label{tbl:ck}
\begin{center}
{
\begin{tabular}{l|l|p{5in}}
\rowcolor{lightgray}
\hline
Metric & Name & Description \\ \hline
amc & average method complexity & Number of Java byte codes\\\hline
avg\_cc & average McCabe & Average McCabe's cyclomatic complexity seen
in class\\\hline
ca & afferent couplings & How many other classes use the specific
class. \\\hline
cam & cohesion amongst classes & \#different
 method parameters types   divided by  
(\#different method parameter types in a   class)*(\#methods).
 \\\hline
cbm &coupling between methods &  Total number of new/redefined methods
to which all the inherited methods are coupled\\\hline
cbo & coupling between objects & Increased when the methods of one
class access services of another.\\\hline
ce & efferent couplings & How many other classes is used by the
specific class. \\\hline
dam & data access & Ratio of  private (protected)
attributes to   total   attributes\\\hline
dit & depth of inheritance tree &  It's defined as the maximum length from the node to the root of the tree\\\hline
ic & inheritance coupling &  Number of parent classes to which a given
class is coupled (includes counts of methods and variables inherited)
\\\hline
lcom & lack of cohesion in methods &Number of pairs of methods that do
not share a reference to an instance variable.\\\hline
locm3 & another lack of cohesion measure &  $\mathit{m},\mathit{a}$ count  the 
$\mathit{methods},\mathit{attributes}$
in a class.  $\mu(a)$  is the number of methods accessing an
attribute.
$\mathit{lcom3}=((\frac{1}{a} \sum_j^a \mu(a_j)) - m)/ (1-m)$.
\\\hline
loc & lines of code & Total lines of code in this file or package.\\\hline
max\_cc & Maximum McCabe & maximum McCabe's cyclomatic complexity seen
in class\\\hline
mfa & functional abstraction & Number of methods inherited by a class
plus number of methods accessible by member methods of the
class\\\hline
moa &  aggregation &  Count of the number of data declarations (class
fields) whose types are user defined classes\\\hline
noc &  number of children & Number of direct descendants (subclasses) for each class\\\hline
npm & number of public methods & npm metric simply counts all the methods in a class that are declared as public. \\\hline
rfc & response for a class &Number of  methods invoked in response to
a message to the object.\\\hline
wmc & weighted methods per class & A class with more member functions than its peers is considered to be more complex and therefore more error prone\\\hline
defect & defect & Boolean: where defects found in post-release bug-tracking systems. \\\hline
\end{tabular}
}
\end{center}

\end{table*}
The rest of this paper is structured as follows.
\tion{background}  introduces the SE case studies explored in this paper (defect prediction) as well as different
approaches and evaluation criteria.
Secondly, it discusses the many sources of variability inherent in software quality predictors.  In summary, between the raw data and the conclusions there are so many choices, some of which are stochastic (e.g., the random number generators that control test suite selection). All these choices introduce $\epsilon$,  a degree of uncertainty in the conclusions.  \tion{ep}
discusses $\epsilon$-domination  for software quality
prediction and proposes DART, a straightforward ensemble method  that can quickly
sample a results space that divides
into a few $\epsilon$-sized regions.
\tion{exp} describes the experimental details in this study.
\tion{results} checks our conjecture.
It will be seen that DART dramatically out-performs
state-of-the-art defect prediction algorithms,
hyper-parameter tuning algorithms, and data pre-processors. 

Based on these results,
we will argue in the conclusion that it is time to consider a fundamentally different approach to software quality prediction.
Perhaps it is time to stop fretting about the numerous
options available for selecting 
data pre-processing methods or machine learning algorithms, then configuring their controlling parameters.
The results of this paper suggest that most of those decisions
are superfluous since so many methods result
in the same output. Accordingly, we recommend doing something like DART; i.e. {\em first} reason about the results
space {\em before}  selecting an appropriate data mining
technology.

One caveat on all these results is that paper has explored
only one domain; i.e., software  defect prediction. 
As to other domains, 
in as-yet-unpublished experiments, we have initial results
suggesting that this simplification might also work elsewhere (e.g., text mining of programmer comments in Stackoverflow; for predicting
Github issue close time; and for detecting programming bad smells). While those results are promising, they are still preliminary.

That said,  the success of this simplification method for defect prediction begs the question: how many other domains in software quality prediction can also be radically simplified? This will be a fruitful
 direction for future work. 
 
 Note that all the data and scripts used in this study are freely available online for use by other researchers\footnote{URL blinded for review.}.




\section{Background and Related Work} \label{sect:background}
 
\subsection{Why Study Simplification?}

In this section, we argue  it is important
to study methods for simplifying quality predictors.
 
We study simplicity since it is very useful to replace $N$ methods with $M \ll N$ methods, especially when the results from the many
are no better than the few.
A bewildering array of new methods for software quality prediction are reported each year (some of which rely on intimidatingly complex mathematical methods) such as deep belief net learning~\cite{wang2016automatically}, spectral-based clustering~\cite{zhang2016cross}, and n-gram language models~\cite{ray2016naturalness}. 
Ghotra et al. list dozens of different data mining algorithms that might be used
for defect predictors~\cite{ghotra2015revisiting}. Fu and Menzies argue that these algorithms
might 
require extensive tuning~\cite{fu2016tuning}. There are many ways to implement that tuning, some of
which are very slow~\cite{tantithamthavorn2016automated}.
And if they were not enough, other computationally expensive methods might also be
required to handle issues
like (say) class imbalance~\cite{agrawal2018better}.

Given recent advances in cloud computing, it is possible to find the best
method for a particular data set via a
``shoot out''  between different methods.
For example, Lessmann et al.~\cite{Lessmann08} and 
Ghotra et al.~\cite{ghotra2015revisiting} explored 22 and 32 different learning algorithms (respectively) for software
quality defect prediction.
Such studies may require days to weeks of CPU time to complete~\cite{fu2017easy}. 
But are such complex and time-consuming studies necessary?
\bi
\item
If there exists some way to dramatically
simplify software quality predictors; then those cloud-based resources would be better used for other tasks. 
\item
Also, Lessmann and Ghotra et al.~\cite{Lessmann08,ghotra2015revisiting} report that many defect prediction methods have equivalent performance. 
\item
Further, as
we show here; there are very simple
methods that perform even better than the methods studied by Lessmann and Ghotra et al.
\ei

 \begin{table*}[!t]
     \caption{Classifiers used in this study. Rankings from Ghotra et al.~\cite{ghotra2015revisiting}.}
     \label{tbl:ourlearners}
     \footnotesize
     \begin{tabular}{l|l|p{4.85in}}
     \hline
    \rowcolor{lightgray} 
    {\bf RANK} & {\bf LEARNER} & {\bf NOTES}\\\hline
     1 (best) & Random Forest~(RF) & 
     Generate conclusions using multiple entropy-based decision trees.\\\cline{2-3}
     &  Logistic Regression~(SL) & Map the output of a regression into $0\le n \le 1$; thus enabling using regression for classification (e.g., defective if $n>0.5$).\\\hline
     2 & K Nearest Neighbors~(KNN) &  Classify a new instance by finding ``k'' examples of similar instances.
     Ghortra et al. suggested
     $K=8$.\\\cline{2-3}
     &  \multirow{2}{*}{Naive Bayes~(NB)} &  Classify a new instance by (a)~collecting mean and standard deviations of attributes in old instances of  different classes; (b)~return the class whose attributes are statistically most similar to the new instance.\\\hline
     3 & Decision Trees~(DT) & Recursively
     divide data by selecting attribute splits
     that reduce the entropy of the class distribution.\\\cline{2-3}
     & Expectation Maximization (EM) & This clustering algorithm uses iterative sampling and repair to derive a parametric expression for each class. \\\hline
    
     4 (worst) & Support Vector Machines~(SMO) &
     Map the raw data into a higher-dimensional space where it is easier to distinguish the examples.
     \\\hline
     \end{tabular}
     \vspace{-0.2cm}
 \end{table*}
 
Another reason to study simplification is that studies can reveal the underlying nature of seemingly
complex problems.
In terms of core science, we argue that the better we understand
something, the better we can match tools to SE. Tools
which are poorly matched to task are usually complex and/or
slow to execute. DART seems a better match for the tasks explored
in this paper since it is neither complex nor slow. Hence,
we argue that DART is interesting in terms of its core scientific
contribution to SE quality prediction.

Seeking simpler and/or faster solutions is not just theoretically
interesting. It is also an approach currently in vogue in
contemporary software engineering. Calero and Pattini~\cite{Calero:2015} comments that ``redesign for greater simplicity'' also motivates
much contemporary industrial work. In their survey of modern
SE companies, they find that many current organizational redesigns
are motivated (at least in part) by arguments based on
``sustainability'' (i.e., using fewer resources to achieve results).
According to Calero and Pattini, sustainability is now a new
source of innovation. Managers used sustainability-based redesigns
to explore cost-cutting opportunities. In fact, they say,
sustainability is now viewed by many companies as a mechanism
for gaining a complete advantage over their competitors.
Hence, a manager might ask a programmer to assess methods
like DART as a technique to generate  more interesting
products.

For all these reasons, we assert that it is high
time to  explore how to  simplify software
quality prediction methods.

 \subsection{Why Study Defect Prediction?}
 The particular software quality predictor explored here is software defect prediction.
 This section argues that this is a useful area
 of research, worthy of exploration and simplification.
 
Software developers are smart, but sometimes make mistakes. Hence, it is essential to test software before the deployment ~\cite{orso2014software,barr2015oracle,yoo2012regression, myers2011art}. 
 Testing is an expensive process. Software assessment budgets are finite while assessment effectiveness increases exponentially with assessment effort~\cite{fu2016tuning}. Therefore, the standard practice is to apply the best available methods on code sections that seem most critical and bug-prone.

Many researchers find that the software bugs are not evenly distributed across the project~\cite{hamill2009common,koru2009investigation, ostrand2004bugs,misirli2011ai}. 
Ostrand et al.~\cite{ostrand2004bugs} studied AT\&T software projects and found 
that they can find 70\% to 90\% bugs in 
first 20\% of the total files in the projects after sorting according
to the file size. Hamill et al.~\cite{hamill2009common} investigated
the common trends in software fault and they reported that 80\% of the faults happened in 20\%
files in the GCC project. Based on these findings on software defect distribution, 
a smart way to perform software testing is to allocate most assessment budgets to
the more defect-prone parts in software projects. Software defect predictors are 
such a strategy, which is to explore the software project and sample the most defect-prone
files/modules/commits. Software defect predictors are never 100\% correct, 
they can be used to suggest where to focus more expensive methods.

Software defect predictors have been proven useful in many industrial settings. 
Misirli et al.~\cite{misirli2011ai} built a  defect prediction model based on Naive Bayes classier for a telecommunications company.
Their results show that defect predictors can predict 87 percent of code defects, decrease inspection efforts by 72 percent, and hence reduce post-release defects by 44 percent; Kim et al.~\cite{kim2015remi} applied defect prediction model, REMI, to API
development process at Samsung Electronics.They reported that REMI predicted the bug-prone APIs with reasonable accuracy~(0.681 F1 score)
and reduced the resources required for executing test cases.

Software defect predictors not only save labor compared with traditional manual methods, but they are also competitive with certain automatic methods.  
A recent study at ICSE'14, Rahman et al. ~\cite{rahman2014comparing} compared (a) static code analysis tools FindBugs, Jlint, and PMD and (b) static code defect predictors (which they called ``statistical defect prediction'') built using logistic  regression.
They found no significant differences in the cost-effectiveness of these approaches.  Given this equivalence, it is significant to
note that static code defect prediction can be quickly adapted to new languages by building lightweight parsers that find in-
formation  like  Table ~\ref{tbl:ck}.  The same is not true for static code analyzers - these need extensive modification before they can be used in new languages.

\subsection{Different Defect Prediction Approaches}\label{sect:diff}
Over the past decade, defect prediction has attracted many attentions from the software research community.  There are many different types of defect predictors
according to the metrics used for building models:

\bi
\item
Module-level based
defect predictors, which use the complexity of software project,
like McCabe metrics, Halstead's effort metrics and CK object-oriented code metrics~\cite{chidamber1994metrics,kafura1987use,mccabe1976complexity}
of Table~\ref{tbl:ck}.
\item
Just-in-time~(JIT) defect prediction
on change level, which utilizes the change metrics collected from the software code~\cite{kamei2013large, kim2008classifying,fu2017unsupervised, mockus2000predicting, yang2016effort}.
\item The first two points represent much of the work in this area.
For completeness, we add there are numerous other kinds of defect predictors based on many
and varied other methods.
For example, Ray et al. propose a defect prediction method using n-gram language models~\cite{ray2016naturalness}. Other work argues that process metrics are more important than the product metrics mentioned in the last two points~\cite{rahman2013and}.
\ei


\begin{figure}[!htp]
\includegraphics[width=0.4\textwidth]{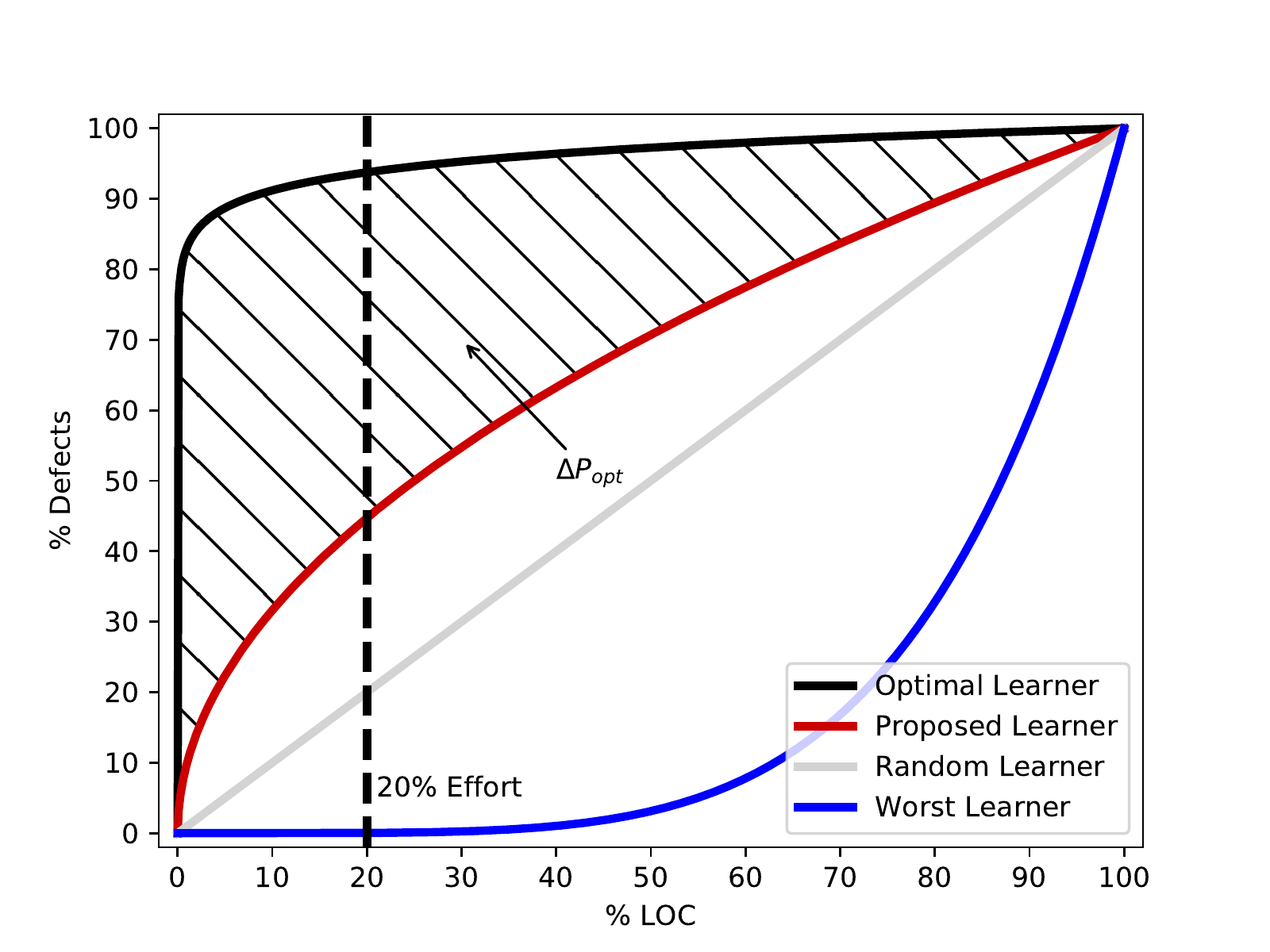}
\caption{Illustration of $P_{opt}$ metric.}
\label{fig:p_opt_demo}
\end{figure}

\begin{table}[!htp]
    \caption {32 defect predictors clustered by their performance rank by Ghotra et al. (using  a Scott-Knot statistical test)~\cite{ghotra2015revisiting}.}
    \label{tbl:learners}
    \footnotesize
    \begin{tabularx}{\linewidth}{l|p{2.7in}}
        \hline
        \rowcolor{lightgray} 
        \centering
        Rank & Classification algorithm \\ \hline
        1  (best) & Rsub+J48, SL, Rsub+SL,  Bag+SL, LMT, RF+SL, Bag+LMT, Rsub+LMT, RF+LMT, RF+J48     \\ \hline
        2 & RBFs, Bag+J48, Ad+SL,  KNN, RF+NB, Ad+LMT, NB, Rsub+NB, Bag+NB   \\\hline
        3 & Ripper, J48, Ad+NB,  Bag+SMO, EM, Ad+SMO, Ad+J48, \\
            & K-means \\\hline
        4  (worst)& RF+SMO, Rsub+SMO,  SMO,  Ridor    \\ \hline 
    \end{tabularx}
    \vspace{-1ex}
\end{table}

Defect prediction models can be built via a variety of   machine learning algorithms such as Decision Tree, Random Forests, SVM, Naive Bayes and Logistic Regression\cite{khoshgoftaar2001modeling,khoshgoftaar2003software,khoshgoftaar2000balancing,menzies2007data,lessmann2008benchmarking,hall2012systematic}. 
Ghotra et al.~\cite{ghotra2015revisiting} compared various classifiers for defect prediction
(for notes on a sample of
those classifiers,
see Table~\ref{tbl:ourlearners}).
According to
their study, the prediction performances of classifiers 
group into the four clusters of   Table~\ref{tbl:learners}.
One advantage of this result is that, to sample across space of
prior defect prediction work
(including some state-of-the-art methods)
researchers need only select one learner from each group.

To improve the performance of defect predictors, 
Fu et al. ~\cite{fu2016tuning, fu2016differential}  and Tantithamthavorn et al. ~\cite{tantithamthavorn2016automated} recommended improving standard
typical defect predictors, like Random Forests and CART by performing hyper-parameter tuning. 
Results from both research groups confirm that hyper-parameter tuning can dramatically supercharge
the defect predictors.
In other tuning work,
Agrawal et al.~\cite{agrawal2018better} argued that  better data is better than a better learner, 
where their results show that defect predictors can be improved a lot by changing
the distribution of defective and non-defective examples seen during
training.

For this paper, to compare the performance of our proposed method, DART, with the state-of-the-art defect
prediction methods, we picked one classification technique at random from each group of Table \ref{tbl:learners}: SL, NB, EM, and SMO.
Furthermore, we adopted techniques from Fu et al. ~\cite{fu2016tuning, fu2016differential} and Agrawal et al.~\cite{agrawal2018better} to investigate how DART performs compared to the improved~(more sophisticated) defect prediction techniques. Note that Agrawal et al.  also selected
different classification techniques from Ghotra et al. study~\cite{ghotra2015revisiting, agrawal2018better}. Hence, using these classifiers, we can also compare our results to the
 experiments of Agrawal et al~\cite{agrawal2018better}.  

\subsection{Evaluation Criteria}\label{sect:eval}

 \begin{figure}[!htp]
\includegraphics[width=0.4\textwidth]{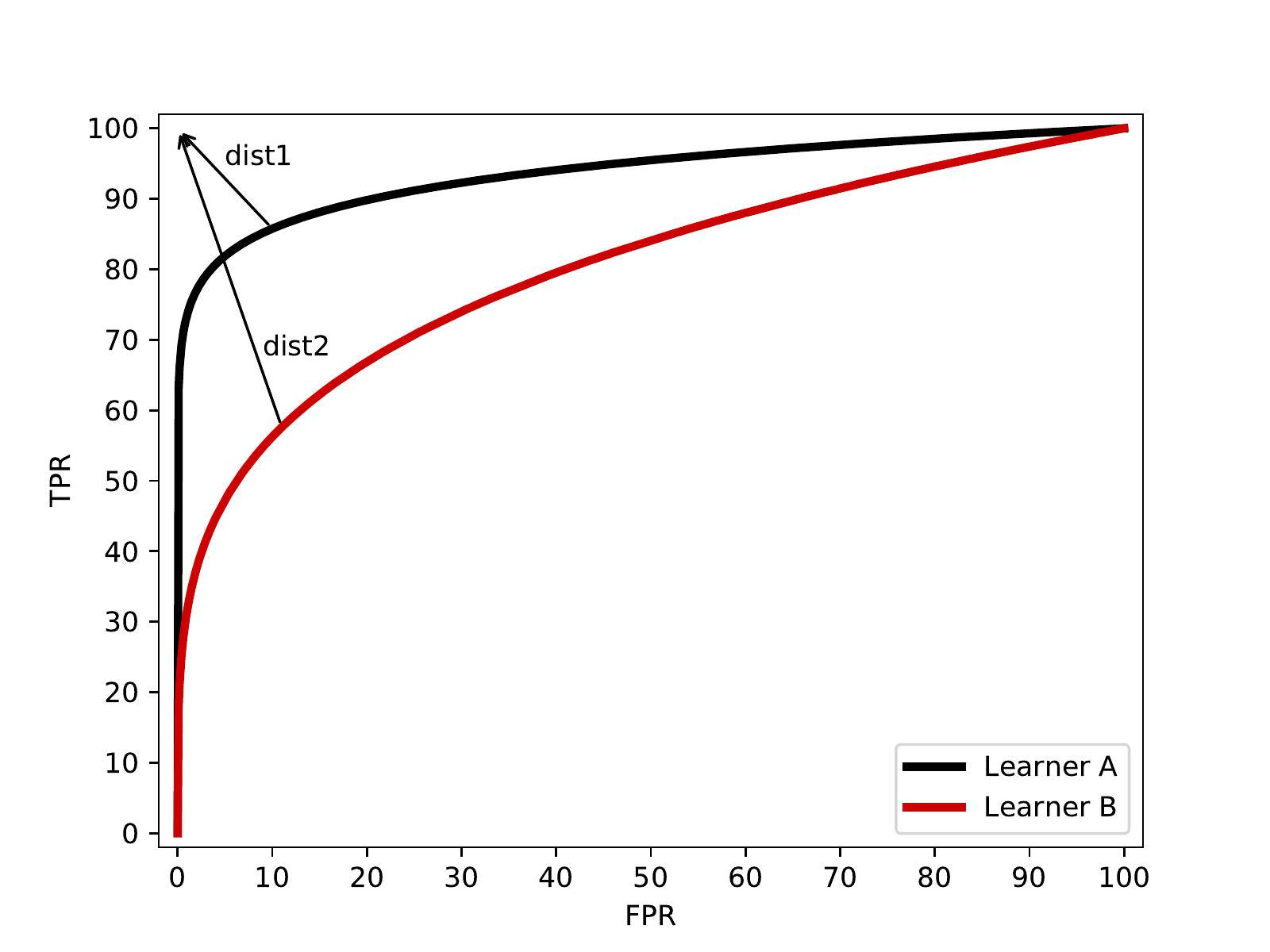}
\caption{Illustration of $dis2heaven$ metric.}
\label{fig:dist2heaven_demo}
\end{figure}





In defect prediction literature, once a learner is executed,  the results must be scored.
{\em 
Recall} measures the percentage of defective modules found by a model generated by the learner.
{\em 
False alarm} reports how many non-defective modules the learner reports as defective.
$P_{\mathit{opt}}$ measures how much code some secondary quality assurance method would have to perform {\em after} the learner has terminated.


$P_{opt}$ is defined as $1- \Delta_{opt}$, where $\Delta_{opt}$ is the area
between the effort~(code-churn-based) cumulative lift charts of the optimal
learner and the proposed learner (as shown in Figure \ref{fig:p_opt_demo}). To calculate $P_{\mathit{opt}}$, we divide all the code modules into those predicted to be defective ($D$) or not ($N$). Both sets are then sorted in ascending order of lines of code. The two sorted sets are 
then laid out across the 
 x-axis,
with  $D$ before $N$. This layout means that 
  the x-axis extends from 0 to 100\% where lower values of $x$ are predicted to be more defective than $x$ higher values.
 On such a chart, the y-axis shows what percent of the defects would be recalled if we traverse the code sorted that x-axis order. 
According to Kamei et al. and Yang et al. ~\cite{yang2016effort,kamei2013large,monden2013assessing}, $P_{opt}$ should be normalized as follows:
\begin{eqnarray} 
P_{opt}(m) = 1- \frac{S(optimal)-S(m)}{S(optimal)-S(worst)} \\\nonumber
\end{eqnarray}
 \begin{wrapfigure}{r}{1.8in}
\centerline{ \includegraphics[width=1.8in]{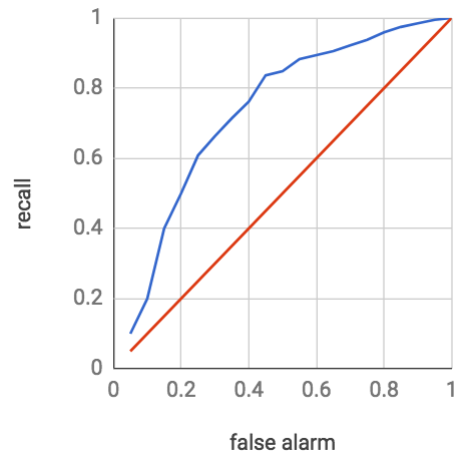}}
 \caption{Results space when  recall is greater than false alarms (see blue curve).}\label{fig:pdpf}
 \end{wrapfigure}
\noindent
where $S(optimal)$, $S(m)$ and $S(worst)$ represent the area of curve under the optimal learner, proposed learner, and worst learner, respectively.
Note that the worst model is built by sorting all the changes according to the actual defect density in ascending order. 
For any learner, it performs better than random predictor only if the ${P_{opt}}$ is greater than 0.5.
 
Note that these measures are closely inter-connected.
 Recall appears as the dependent variable of $P_{\mathit{opt}}$. 
 Also,
 false alarms result in  flat  regions of the $P_{\mathit{opt}}$ curve.
 Further, for useful learners, recall is greater than the false alarm. Such learners
   have the characteristic shape of \fig{pdpf}.

 In the following, we will assess results using $P_{\mathit{opt}}$
 and another measure ``distance to heaven'' (denoted {\em dis2heaven})
 that computes the distance of some recall, false alarm pair to
 the ideal ``heaven'' point of recall$=1$ and false alarm$=0$ as shown in Figure~\ref{fig:dist2heaven_demo}.
 This measure is defined in Equation~\ref{eq:recall}:
 
\begin{equation}\label{eq:recall}
\mathit{dis2heaven} =  
\frac{\sqrt{{(1-\mathit{recall})}^2 + {(\mathit{false\;alarm}})^2}}{\sqrt{2}}
\end{equation} 
The denominator of this equation means that $0 \le \mathit{dis2heaven} \le 1$.
Note that :
\bi
\item For $P_{\mathit{opt}}$, the {\em larger} values are {\em better};
\item For {\em dis2heaven}, the {\em smaller} values are {\em better};
\ei
We use these measures instead of, say, precision or the F1 measure (harmonic mean of precision and recall) since Menzies et al.~\cite{Menzies:2007prec} warn that precision can be very unstable for SE   data (where the class distributions may be highly unbalanced).

\subsection{Sources of $\epsilon$ Uncertainty}\label{sect:var}
 
 
 This section argues that there is inherent uncertainty in making conclusions via software quality predictors. The rest of this paper exploits that uncertainty to simplify defect prediction.
 
 Given that divergent nature of software projects and software developers, 
 it is to be expected that different researchers find different effects, even from the same data sets~\cite{hosseini2017systematic}. According to Menzies et al.~\cite{menzies2012special},
 the conclusion uncertainty of software quality predictors come from different choices in the training data and many other factors.
 
  {\em Sampling Bias:}   Any data mining algorithm needs to input multiple examples to make its conclusions. The more diverse the input examples, the greater the variance in the conclusions. And software engineering is a very diverse
 discipline: 
  
 \bi
 \item
 The software is built by engineers
 with varying skills and experience.
 \item
 That construction process is performed using
 a wide range of languages and tools. 
 \item
 The completed product is delivered on lots of platforms. 
 \item
 All those languages, tools and platforms keep changing and evolving. 
 \item
 Within one project, the problems
 faced and the purpose served by each software module may be very different (e.g., GUI, database, network connections, business logic, etc.). 
 \item
 Within the career of one developer, the problem  domain,  goal, and stakeholders of their software
 can change dramatically from one project to the next.  
 \ei
 {\it Pre-processing}: Real world data usually requires some clean-up before it can be used effectively by data mining algorithms.   There are many ways to transform the data favorably. The numeric data may be discretized into smaller bins. Discretization can greatly affect the results of the learning: since there may be ways to implement discretization~\cite{fayyad1993multi}; Feature selection is sometimes useful to prune uncorrelated features to the target variable~\cite{chen2005feature}. On the other hand, it can be helpful to prune data points that are very noisy or are outliers~\cite{kocaguneli2010use}. The effects of pre-processing can be quite dramatic. For example, Agrawal et al.~\cite{agrawal2018better} report that their pre-processing technique~(SMOTUNED) increased AUC and recall by $60\%$ and $20\%$, respectively. Note that the choices made during pre-processing can introduce some variability in the final results.
    
 {\it Stochastic algorithms}: Numerous methods in software quality predictors employ stochastic algorithms that use random number generators.
    For example, the key to scalability is usually (a)~build a model on the randomly selected small part of the data then (b)~see how well that works over the rest of the data~\cite{Sculley:2010}. Also, when evaluating data mining algorithms,
    it is standard practice to divide the data randomly into several bins as part of a cross-validation experiment~\cite{Witten:2002}. For all these stochastic algorithms, the conclusions are adjusted, to some extent, by the random numbers used in the processing.

 Many methods have been proposed to reduce the above uncertainties such as
 feature selection to remove spurious outliers~\cite{menzies2007data}, application of
 background knowledge to constrain model generation~\cite{fenton2012risk}, optimizers to tune model parameters to reduce uncertainty~\cite{Agrawal18stab}.
 Despite this, some uncertainty $\epsilon$ usually remains (see an example, next section). 
 
 We conjecture that, for all the above reasons,
uncertainty is an inherent property of software quality prediction. If so, the question becomes, ``what to do with that uncertainty?''.
 The starting point for this paper was the following speculation:
 \textbf{\em{ Instead of striving to make $\epsilon=0$, use $\epsilon>0$ as a tool for simplifying software quality predictors}}. The next section describes such a tool.
 
 \section{$\epsilon$-Domination}\label{sect:ep}
 
From the above, we assert that software quality predictors result collected
on the same data will
vary by some amount $\epsilon$.
As mentioned in the introduction,
Deb's principle of $\epsilon$-dominance~\cite{deb2005evaluating} 
states that if there exists
some  $\epsilon$ value below
which is useless or impossible to distinguish results, then
it is superfluous to explore anything less than $\epsilon$.

\begin{wrapfigure}{r}{1.2in}
\includegraphics[width=1.2in]{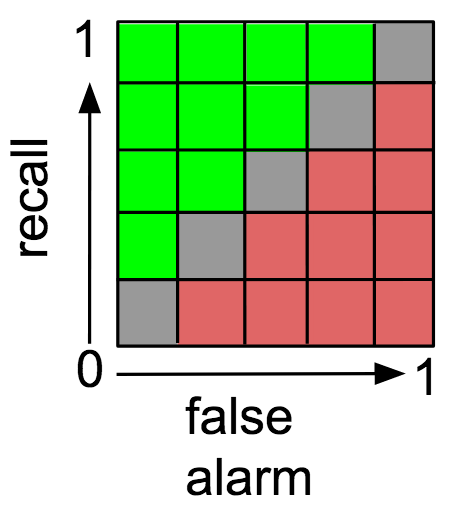}
\caption{Grids in results space.}\label{fig:resultspace}
\end{wrapfigure}
Note that $\epsilon$ effectively clusters the space of possible results.
For example, consider the result space defined by recall $r$ and false alarms $f$.
Both these measures have the  range \mbox{$0 \le r \le 1$} and
 \mbox{$0 \le f \le 1$}.
 If $\epsilon=0.2$,  then the results space of possible recalls and false alarms divides into the 5*5 grid of \fig{resultspace}.  
 
 \fig{pdpf} showed that the results from useful learners have a characteristic shape where recall is greater than false alarms. That is, in \fig{resultspace},
 such results avoid
 the red regions of that grid (where false alarms are higher than recall)
 and the gray regions (also called the ``no-information'' region where recall
 is the same as a false alarm). 
 
 This  means that when  \mbox{$\epsilon=0.2$}, then
(a)~{\em recall-vs-false alarm} results space is effectively just the ten green cells of \fig{resultspace};
and (b)~``many roads lead to Rome'' (i.e., if the results of 100 learners were places on this grid, then there could never be more than 10 groups of results).
 
\begin{figure}[!t]
\centerline{\includegraphics[width=2.8in]{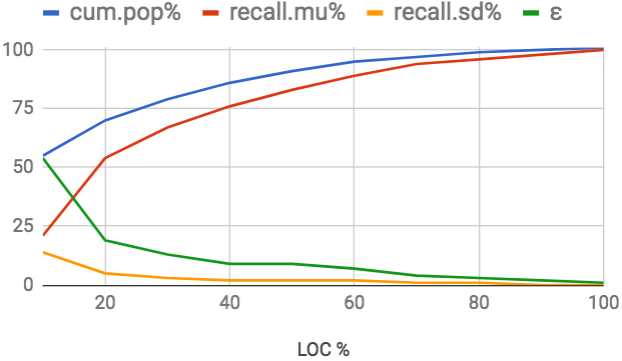}}
\caption{$\epsilon$ can vary across results space.
100 experiments with LUCENE results using 90\% of the data, for training,
and 10\%, for testing. The x-axis sorts the code base, first according to predicted defective or not, then second on lines of code.
The blue line shows the distribution of the 100 results across this space. For example, 70\% of the results predict defect for up to
  20\% of the code (see the blue curve).
The y-axis of this figures shows mean recall (in red); the standard
deviation $sigma$ of the recall (in yellow); and $\epsilon$ is defined as per standard t-tests that says at the 95\% confidence level, two distributions differ
when they are   $\epsilon=2*1.96*\sigma$ apart.
(in green). }\label{fig:lucene}
\end{figure}

    It turns out that real-world results spaces are more complicated than shown in \fig{resultspace}. For example, consider the results space of \fig{lucene}. In this figure, 100 times, a defect predictor was built for LUCENE, an open-source Java text search engine.
  Random Forests was used to build the defect predictor 
  using 90\% of the data,
 then tested on the remaining 10\% (Random Forests are a multi-tree classifier, widely used in defect prediction; see Table~\ref{tbl:ourlearners}).
 
 To compute $\epsilon$ in this results space, we divide the x-axis into divisions of 0.1 and report the standard deviation $\sigma$ of recall in each division. 
 For the moment, we use  a simple
 t-test to infer the   separation required
 to distinguish two results within \fig{lucene} (later in this section, we will dispense with that assumption).
 This means that 
   $\epsilon=2*1.96*\sigma$ which is the
   the range required to be 95\% confident that two distributions are different~\cite{Witten:2005}.

The main result of \fig{lucene} is       
  that $\epsilon$ is often very large.
The blue curve shows  that 70\% of the results occur in the region
\mbox{$0\le\mathit{LOC}\le 0.2$}.
At $\mathit{LOC}=0.2$, $\epsilon= 0.2$; i.e.,   most of our
models   have an $\epsilon$ of 0.2 or higher.
Note that   learners with $\epsilon \ge 0.2$ divide into
the 25 cells, or less, of  \fig{resultspace}.
More specifically, it  means that most of the results of 100 learners applied to LUCENE would have  
  statistically indistinguishable results.

From an analytic perspective, there are some limitations with the above analysis. Firstly,  the threshold of $2*1.96*\sigma$ is a simplistic
measure of statistically significantly different results. It makes many assumptions that may not hold for SE data; e.g., that the
data conforms to a parametric Gaussian distribution and that the variance
of the two distributions is the same.

Secondly,   as shown by the green curve of \fig{lucene}, $\epsilon$ is not uniform across this result space. 
One reason for the lack of uniformity is that the results
generated from 100 samples of the LUCENE data do not fall evenly across space:
70\% of the 100 learned models
 fall far left of \fig{lucene} (up to  20\% of the code-- see the blue curve).
 This high variance  means that 
 we cannot reason about the results of space just via, e.g.,  some trite summary of the entire results space as a mean $\epsilon$ value.
 

When analytic methods fail,  sampling can use instead.
Rather than using $\epsilon$ analytically, we instead use it to define
a  sampling method of the results space. That system, called DART is described in \fig{dart}.
The algorithms work by DART-ing around results space, a couple of times.
Note that if the results exhibit a large $\epsilon$ properties, then
these few samples would be enough to
cover  the ten green cells of \fig{resultspace}. Also, the results from such DART-ing around
should perform as well as anything else,  including the three state-of-the-art systems listed in the introduction.

To operationalize DART, we wrote some Python code
based on the Fast-and-frugal tree (FFT)
R-package from Phillips et al.~\cite{phillips2017fftrees}. While this is not the only way to operationalize DART, it worked so well
for this paper; we were not motivated to try alternatives.
An FFT is a binary tree where, at each level, there is one exit node predicting
either for ``true'' for target class or ``false''. Also, at the bottom of the tree, there are two leaves exiting to ``true'' and ``false''.
For example, from the Log4j dataset of Table~\ref{tbl:data}, one tree predicting for software defects is shown in Figure.~\ref{fft}. Note that this tree has decided to exit towards the target class at lines 2, 3, 4 and otherwise on lines 1, 5.

\begin{figure}[!t]
\small \begin{tabular}{|p{.95\linewidth}|}\hline
INPUT: 
\bi
\item A dataset, such as Table~\ref{tbl:data};
\item A goal predicate $p$; e.g., ${P_{\mathit{opt}}}$ or  $\mathit{dis2Heaven}$;
\item $M,N$=  number of models, number of ranges used per model
\ei
OUTPUT:
\bi\item Score of the best model when applied to data not used for training.
\ei 
PROCEDURE:
\bi
\item Separate the data into train and test;
\item On the train data, build an ensemble and select the best:
\bi
\item   For $i=1$ to $M$  do
\be
\item Divide numeric attributes into ranges; 
\item Find $N$ {\em extreme} ranges  that score highest and lowest on $p$;  
\item  Combine some the extreme ranges into   model $i$;
\item Score $i$   using $p$;
\item Keep the best scoring model.
\ee
\ei
\item On the test data:
\bi
\item Return the $p$ score of  the best scoring model.
\ei
\ei 
NOTES:
\bi
\item
For training step (2), we use  {\em extreme} ranges in order to maximize the spread of the darts around the results space.
\item
To  keep this simple,  the  discretizer  used in   training step (1)  just divides the
numeric data on its median value.
\ei\\\hline
\end{tabular}
\caption{DART: an ensemble algorithm to   sample results space,
$M$  number of times.}\label{fig:dart}
\end{figure}

\begin{figure}[!t]
\small \begin{tabular}{|p{.95\linewidth}|}\hline
1.  if      cob <= 4    \qquad \qquad then false     \\ 
2.  else if rfc > 32    \qquad \ \, then true       \\
3.  else if dam >  0    \qquad \: then true       \\
4.  else if amc < 32.25 \quad then true       \\
5. else false \\\hline
\end{tabular}
\caption{A simple model for software defect prediction}\label{fft}
\end{figure}


 To build one tree, our version of DART discretize numerics by dividing   at   median values; 
 then scores each range using {\em dis2heaven} or $P_{\mathit{opt}}$ according to how well they predict for ``true'' or ``false''
 (this finds the  {\em extreme} ranges seen in DART's training step (2)).
 Next we built one   level of the tree by (a)~picking the exit class then (b)~adding in the range that best predicts for that class.
 The other levels are build recursively using the data not selected by that range.  
 Given a tree of depth $d$, there are  two choices at each level about whether or not to exit to ``true'' or ``false''.
 Hence, for trees of depth $d$, there are   $M=2^d$ possible trees. Each such tree is one ``dart'' into
 results space.
 
 To throw several darts at results space,
 DART builds an  ensemble of 16 trees, we use  depth $d=4$.
 This number was selected since   \fig{resultspace} had   ten green cells. Hence: $d=3$ would generate $2^3=8$ trees
 which would not be enough to cover results space;
 $d=5$  would generate $2^5=32$ trees which would
 be excessive for results like \fig{resultspace}.
 Note that, when using this approach, the number of extreme ranges used in the models is the same as the depth of the tree $N=d=4$.
 As per \fig{dart}, on the training data shown in Table~\ref{tbl:data}, we built 16 trees, then selected the best one
 to be used for testing.

\section{Experimental SETUP}\label{sect:exp}

Recalling \tion{eval}, the evaluation criteria used in this study was {\em dis2heaven} or $P_{\mathit{opt}}$. Note that this criteria also echoes the criteria seen in prior work~\cite{fu2017unsupervised, kamei2013large, yang2016effort}. The rest of this section discusses our other experimental details.

\subsection{Research Questions}
 To compare with three established defect prediction methods, we use all machine learning implementations from Scikit-learn package
and tools released by Fu et al.~\cite{fu2016tuning} and Agrawal et al.~\cite{chen2018multi}. In this study, we set three research questions:

\textbf{RQ1:} {\it Do established learners sample results space better than a few DARTs?} This questions compares DART
against the sample of defect prediction algorithms surveyed by
Ghotra et al. at ICSE'15~\cite{ghotra2015revisiting}.

\textbf{RQ2:} {\it Do goal-savvy learners sample results space better than a few DARTs?} This question address a potential problem with the {\bf RQ1} analysis.   DART uses the goal function when it trains its models. Hence, this might  give DART an unfair advantage
compared to other learners in Table~\ref{tbl:ghotraout}. Therefore,
in {\bf RQ2}, we compare DART to {\em goal-savvy}
hyper-parameter optimizers~\cite{fu2016tuning}  that make extensive use
of the goal function
as they tune learner parameters.

\textbf{RQ3:} {\it Do data-savvy learners sample results space better than a few DARTs?} 
Agrawal et al.~\cite{chen2018multi} argues that selecting and/or tuning
data miners is less useful that repairing problems with the training
data. To test that, this research question compares DART
agains the {\em data-savvy} methods developed by Agrawal et al.

\subsection{Datasets}
To compare the DART ensemble method against alternate approaches, we used data from SEACRAFT repository (tiny.cc/seacraft),  shown in Table~\ref{tbl:data}
(for details on the contents of those data sets, see Table~\ref{tbl:ck}).
This data was selected for two reasons:
\bi
\item The data is available for multiple versions of the same software. This means we can  ensure that our learners are trained on past data and tested on future data.
\item It is very similar, or identical, to the data used  in prior work against which we will compare our new approach
~\cite{ghotra2015revisiting,fu2016tuning,agrawal2018better}.
\ei

When applying data mining algorithms to build predictive models,
one important principle is not to test on the data used in training. 
There are many ways to design a experiment that satisfies this principle. 
Some of those methods have limitations; e.g., leave-one-out is too slow for
large data sets and cross-validation mixes up older and newer data~
(such that data from the past may be used to test on future data). In
this work, for each project data, we set the latest version of project data
as the testing data and all the older data as the training data.
For example, we use $\mathit{poi1.5, poi2.0, poi2.5}$ data for training predictors,
and the newer data, $\mathit{ poi3.0}$ is left for testing.

 \begin{table}[!t]
\centering

\scriptsize
\caption{Statistics of the studied data sets.}
\label{tbl:data}
\resizebox{0.48\textwidth}{!}{
\begin{tabular}{cllll}
\hline
\rowcolor{lightgray} \multicolumn{1}{c|}{}& \multicolumn{2}{c|}{Training Data}& \multicolumn{2}{c}{Testing Data}\\ \cline{2-5} 
\rowcolor{lightgray}\multicolumn{1}{c|}{\multirow{-2}{*}{Project}} & \multicolumn{1}{c|}{Versions} & \multicolumn{1}{c|}{\% of Defects} & \multicolumn{1}{c|}{Versions} & \multicolumn{1}{l}{\% of Defects} \\ 
\hline

\multicolumn{1}{l}{Poi}& \multicolumn{1}{l}{1.5, 2.0, 2.5}& \multicolumn{1}{l}{426/936 = 46\%}& \multicolumn{1}{c}{3.0}& \multicolumn{1}{l}{281/442 = 64\%} \\ \hline
\multicolumn{1}{l}{Lucene}& \multicolumn{1}{l}{2.0, 2.2}& \multicolumn{1}{l}{235/442 = 53\%}& \multicolumn{1}{c}{2.4}& \multicolumn{1}{l}{ 203/340 = 60\%} \\ \hline
\multicolumn{1}{l}{Camel}& \multicolumn{1}{l}{1.0, 1.2, 1.4}& \multicolumn{1}{l}{374/1819 = 21\%}& \multicolumn{1}{c}{1.6}& \multicolumn{1}{l}{ 188/965 = 19\%
} \\ \hline
\multicolumn{1}{l}{Log4j}& \multicolumn{1}{l}{1.0, 1.1}& \multicolumn{1}{l}{71/244 = 29\%}& \multicolumn{1}{c}{1.2}& \multicolumn{1}{l}{189/205 = 92\%} \\ \hline
\multicolumn{1}{l}{Xerces}& \multicolumn{1}{l}{1.2, 1.3}& \multicolumn{1}{l}{140/893 = 16\%}& \multicolumn{1}{c}{1.4}& \multicolumn{1}{l}{437/588 = 74\%} \\ \hline
\multicolumn{1}{l}{Velocity}& \multicolumn{1}{l}{1.4, 1.5}& \multicolumn{1}{l}{289/410 = 70\%}& \multicolumn{1}{c}{1.6}& \multicolumn{1}{l}{78/229 = 34\%} \\ \hline
\multicolumn{1}{l}{Xalan}& \multicolumn{1}{l}{2.4, 2.5, 2.6}& \multicolumn{1}{l}{908/2411 = 38\%}& \multicolumn{1}{c}{2.7}& \multicolumn{1}{l}{898/909 = 99\%} \\ \hline
\multicolumn{1}{l}{Ivy}& \multicolumn{1}{l}{1.1, 1.4}& \multicolumn{1}{l}{79/352 = 22\%}& \multicolumn{1}{c}{2.0}& \multicolumn{1}{l}{40/352 = 11\%} \\ \hline
\multicolumn{1}{l}{Synapse}& \multicolumn{1}{l}{1.0, 1.1}& \multicolumn{1}{l}{76/379 = 20\%}& \multicolumn{1}{c}{1.2}& \multicolumn{1}{l}{ 86/256 = 34\%} \\ \hline
\multicolumn{1}{l}{Jedit}& \multicolumn{1}{l}{ \begin{tabular}[c]{@{}c@{}}3.2, 4.0 \\ 4.1, 4.2 \end{tabular}}& \multicolumn{1}{c}{292/1257 = 23\%}& \multicolumn{1}{c}{4.3}& \multicolumn{1}{l}{11/492 = 2\%} \\ \hline
\end{tabular}
}
\end{table}

\begin{table}[t!]
    \caption { DART v.s. state-of-the-art defect predictors from Ghotra et al.~\cite{ghotra2015revisiting} for $\mathit{ dis2heaven}$ and $P_{\mathit{opt}}$.
    Gray cells mark best performances on each project (so DART is most-often best).}
    \label{tbl:ghotraout}
    \centering
    \footnotesize
    \begin{tabular}{c|l|ccccc}
        \rowcolor{lightgray}
        \hline
        Goal &  Data & DART & SL & NB & EM  & SMO \\ \hline
        \multirow{10}{*}{\rotatebox[origin=c]{90}{\parbox[c]{3cm}{{\em dis2heaven: ({\em less} is {\em better})}}}}   &  log4j & \colorbox{gray!30}{23} & 53 & 51 & 56 & 48 \\ 
          &  jedit &\colorbox{gray!30}{31} & 40 & 41 & 34 & 47 \\
          &  lucene & \colorbox{gray!30}{33} & 40 &	44 & 44	& 71 \\
          &  poi & \colorbox{gray!30}{35} & 36 &	57 & 70 & 45 \\
          &  ivy & \colorbox{gray!30}{35} &	50	&40	& 71	& 43 \\
          &  velocity & \colorbox{gray!30}{37} & 61 &	40 &	49 & 60 \\
          &  synapse & 38	& 51	& 39	& \colorbox{gray!30}{34} & 62 \\
          &  xalan &  \colorbox{gray!30}{39} & 55 &	55 &	70	& 68 \\
          &  camel & \colorbox{gray!30}{41} &	60 &	52& 44	&71 \\ 
          &  xerces & \colorbox{gray!30}{42} & 68 & 60	& 50& 69 \\ \hline
 
       \multirow{10}{*}{\rotatebox[origin=c]{90}{\parbox[c]{2cm}{$P_{\mathit{opt}}$: ({\em more} is {\em better})}}}    &  ivy & \colorbox{gray!30}{28} & 17 & 9 & \colorbox{gray!30}{28} & 23 \\ 
          &  jedit & \colorbox{gray!30}{39} & 10 & 9 & 16 & 17  \\   
          &  synapse & \colorbox{gray!30}{43} & 26 & 24 & 22 & 22 \\  
          &  camel & \colorbox{gray!30}{53} & 15 & 17 & 16 & 50 \\   
          &  log4j & \colorbox{gray!30}{56} & 19 & 22 & 16 & 23 \\  
          &  velocity & \colorbox{gray!30}{64} & 64& \colorbox{gray!30}{64} & 24 & 60 \\   
          &  poi & \colorbox{gray!30}{73} & 51 & 19 & 33 & 64 \\  
          &  lucene & \colorbox{gray!30}{81} & 43 & 27 &20 & 80 \\ 
          &  xerces & \colorbox{gray!30}{90} & 4 & 9 & 15 & 48 \\   
          &  xalan & 99 & 11 & 15& \colorbox{gray!30}{100} & 51 \\\hline   
        \end{tabular}
\end{table}
 
\section{Results}\label{sect:results}

 \begin{table}
    \centering
    \caption{DART v.s. tuning Random Forests for $\mathit{ dis2heaven}$ and $P_{\mathit{opt}}$. Gray cells mark best performances on each project. Note that even when DART does not perform best, it usually performs very close to the best.}
    \footnotesize
    \resizebox{0.37\textwidth}{!}{
    \begin{tabular}{r|cc|cc}
    \hline
        \multirow{3}{*}{Data} &
        \multicolumn{2}{c|}{{{\em dis2heaven}}} &  \multicolumn{2}{c}{$P_{opt}$} \\
           &
        \multicolumn{2}{c|}{{{\em (less is better)}}} &  \multicolumn{2}{c}{{\em (more is better)}} \\
        \cline{2-5}
          &DART & Tuning RF & DART & Tuning RF  \\\hline
        ivy & 	\colorbox{gray!30}{35} & 56 & 	\colorbox{gray!30}{28} & 	\colorbox{gray!30}{28}\\ 
        jedit & 	\colorbox{gray!30}{31} & 35 & 	\colorbox{gray!30}{39} & 	\colorbox{gray!30}{39} \\ 
        synapse & 	\colorbox{gray!30}{38} & 57 & 43 & 	\colorbox{gray!30}{48} \\ 
        camel & 	\colorbox{gray!30}{41} & 70 & 53 & 	\colorbox{gray!30}{54} \\ 
        log4j & 	\colorbox{gray!30}{23} & 51 & 	\colorbox{gray!30}{56} & 20 \\ 
        velocity & 	\colorbox{gray!30}{37} & 53 & 64 & 	\colorbox{gray!30}{64} \\ 
        poi & 34.8 & 	\colorbox{gray!30}{27} & 73 & 	\colorbox{gray!30}{74} \\ 
        lucene & 	\colorbox{gray!30}{33} & 35 & 	\colorbox{gray!30}{81} & 80 \\ 
        xerces & 	\colorbox{gray!30}{42} & 70 & 90 & 	\colorbox{gray!30}{94} \\ 
        xalan & 38.7 & 	\colorbox{gray!30}{36} & 	\colorbox{gray!30}{99} & 	\colorbox{gray!30}{99} \\ \hline 
    \end{tabular}
    }
    \label{tab:rq2}
\end{table}
 
\subsection{RQ1: Do established learners sample results space better than a few DARTs?}

 In order to compare our approach to established norms in software quality predictors,
 we used the Ghotra et al. study from ICSE'15~\cite{ghotra2015revisiting}. Recall that this study was a comparison
 of the the 32 learners shown in 
Table~\ref{tbl:learners}. The performance of those learners clustered into four groups, from which we selected four representative learners
(see the discussion   in \tion{diff}): SL, NB, EM, SMO.

For this comparison, DART and the learners from Ghotra et al.  were all trained/tested on the same versions 
shown in Table~\ref{tbl:data}. The resulting performance scores are shown in Table~\ref{tbl:ghotraout}. Note
that:
\bi
\item
DART performed as well, or better, than the sample of Ghortra et al learners in 18/20 experiments.
\item When DART failed to produce best performance, it came very close to the best (e.g., for 
Xalan's  $P_{\mathit{opt}}$ results, DART scored 99\% while the best was 100\%).
\item
When DART performed best, it often did so by a very wide margin.  For example,
for Log4j's {\em dis2heaven} score, DART's score was 23\% and the best value of the other learners was 51\%; i.e., worse
by more than a factor of two. For another example, for Log4j's $P_{\mathit{opt}}$ score, DART's score was more than twice better
than the scores of any other learner.
\ei
From these results, we assert that DART out-performs the established state-of-the-art defect predictors recommended by Ghortra el al.~\cite{ghotra2015revisiting} on the data sets
of Table~\ref{tbl:data}.

\begin{table*}[!t]
\caption {List of parameters tuned in this paper.}
\centering
    \resizebox{\textwidth}{!}{
    \begin{tabular}{c|c|c|c|l}
    \hline
    \rowcolor{lightgray}
    
    \begin{tabular}[c]{@{}c@{}}Tuning Object\end{tabular} & Parameters & Default &\begin{tabular}[c]{@{}c@{}}Tuning\\ Range\end{tabular}&\multicolumn{1}{c}{Description} \\ \hline
     \multirow{5}{*}{\begin{tabular}[c]{@{}c@{}}Random \\ Forests\end{tabular}} 
     & threshold & 0.5 & [0.01,1] & The value to determine defective or not. \\\cline{2-5} 
     & max\_feature & None &[0.01,1]& The number of features to consider when looking for the best split. \\ \cline{2-5} 
     & max\_leaf\_nodes & None &[1,50]& Grow trees with max\_leaf\_nodes in best-first fashion. \\ \cline{2-5} 
     & min\_sample\_split & 2 &[2,20]& The minimum number of samples required to split an internal node. \\ \cline{2-5} 
     & min\_samples\_leaf & 1 &[1,20]&The minimum number of samples required to be at a leaf node. \\ \cline{2-5} 
     & n\_estimators & 100 & [50,150]&The number of trees in the forest.\\ \hline 
     \multirow{3}{*}{\begin{tabular}{c} SMOTE \end{tabular}}
     & k & 5 &[1,20] & Number of neighbors \\ \cline{2-5}
     & m & $50\%$& 50, 100, 200, 400 & Number of synthetic examples to create. Expressed as a percent of final training data. \\\cline{2-5}
     & r &  2     & [0.1, 5] & Power parameter for the Minkowski distance metric. \\\hline
     
    \end{tabular}
    }
    \label{tbl:parameters}
\end{table*}

\begin{figure}[!t]
\small \begin{tabular}{|p{.95\linewidth}|}\hline
INPUT: 
\bi
\item A dataset, such as Table~\ref{tbl:data};
\item A tuning goal $G$; e.g.,  ${P_{\mathit{opt}}}$ or  $\mathit{dis2Heaven}$;
\item DE parameters: $\mathit{np}=10$, $\mathit{f}=0.75$, $\mathit{cr}=0.3$, $\mathit{life}=5$
\ei
OUTPUT:
\bi\item Best tunings for learners~(e.g., RF) found by DE
\ei 
PROCEDURE:
\bi
\item Separate the data into $\mathit{train}$ and $\mathit{tune}$;
\item Generate $np$ tunings as the initial population;
\item Score each tuning $\mathit{pop_i}$ in the population with goal $G$; 

\item  For $i=1$ to $np$  do
\be
\item Generate a mutant $m$ built by extrapolating between three other members of population $a$, $b$, $c$ at probability $\mathit{cr}$. For each decision $m_k \in m$:
     \bi
      \item  $m_k= a_k + f*(b_k-c_k)$ (continuous values).
      \item  $m_k= a_k \vee ~ ( b_k \vee c_k)$ (discrete values).
     \ei
\item Build a learner with parameters $m$ and train data;
\item Score $m$ on tune data using $G$;
\item Replace $\mathit{pop_i}$ with $m$ if $m$ is preferred;
\ee
 \item Repeat the last step until run out of $\mathit{life}$ or could not find better tunings;
\item Return the best tuning $\mathit{pop_i}$  of the last population as the final result.
\ei 
\\\hline
\end{tabular}
    \caption{TUNER is an evolutionary optimization algorithm based  on Storn's differential evolution algorithm~\cite{storn1997differential, fu2017easy}. }    \label{fig:DE}
\end{figure}

\subsection{RQ2: Do goal-savvy learners sample results space better than a few DARTs?}

One counter argument to the conclusions of the {\bf RQ1} is that it may not be fair to compare DART against
standard data mining algorithms using their off-the-shelf parameter tunings.  DART makes extensive use of the goal function $p$
(i.e., $P_{\mathit{opt}}$ and {\em dis2heaven}) at three points in its algorithm:
\bi
\item Once when assessing individual ranges;
\item Once again when assessing trees built from those ranges;
\item A third time when assessing the best tree on the test set.
\ei
All the  other learners in Table~\ref{tbl:ghotraout}  use $p$ only once (when their final model was assessed) but never while they
build their models. That is, DART is ``goal-savvy''
while all the other
learners explored in {\bf RQ1} were not.
Perhaps this gave DART an unfair advantage?

To address this issue, we turned to the state-of-the-art in hyper-parameter optimization for defect prediction. 
In 2016, Fu et al.
presented in the IST journal~\cite{fu2016tuning} an extensive study where an optimizer tuned the control parameters
of various learners applied to software quality defect prediction. That study used the goal function $p$ to guide
their selection of control parameters; i.e., unlike the Table~\ref{tbl:ghotraout} results, this learning method
is ``goal-savvy'' in the sense that it was allowed to  reflect on the goal  during model generation.

For RQ2, we   compare the performance of DART with 
goal-savvy tuning Random Forests.
We use RandomForests since they where
recommended by Ghotra et al.~\cite{ghotra2015revisiting} and prior work hyper-parameter
 tuning for defect prediction by   Fu et al.~\cite{fu2016tuning}.
In this hyper-parameter tuning experiment, for each project data,
we randomly split the original training data (e.g, combine $\mathit{poi1.5, poi2.0}$ and $poi2.5$ as the original training data) into $80\%$ and $20\%$ as new training 
data and tuning data, respectively. As recommended by Fu et al.~\cite{fu2016tuning}, we use the TUNER algorithm
of Figure~\ref{fig:DE} to select
the parameters of Random Forests
(for a list of those parameters, see Table~\ref{tbl:parameters}).
TUNER iterates
until it 
runs out of tuning resources (i.e., a given tuning budget) or it 
canot not find any better hyper-parameters. Finally, we use the current best
parameters as the  best hyper-parameters to train Random Forests
with the new training data. 

Since different data split might have
an impact on predictor's performance, we repeat tuning+testing process
30 times, each time with different random seed and return the median
values of 30 runs as the result of tuning Random Forests experiment on
each project data. Since we have two different goals: minimize $\mathit{dis2heaven}$
and maximize $P_\mathit{opt}$ metrics, we run two different experiments (so 60 repeats in all).

For comparison purposes, DART built its ensembles on the original training data (e.g, combine $\mathit{poi1.5, poi2.0}$ and $poi2.5$ as the original training data) and selected the best tree in terms of $p$~(i.e., $P_{opt}$ or $\mathit{dis2heaven}$). The best tree  was then tested on testing data~(e.g., $poi3.0$). Table~\ref{tab:rq2} shows the results of DART versus Random Forests, where the latter was tuned for {\em dis2heaven} or $P_{opt}$.
Note that in this experiment, DART was not tuned. Rather, it just used its default settings of:
\bi
\item Discretizing using median splits for numeric attributes;
\item Building trees of depth $d=4$, which means ensembles of size $2^d=16$;
\item At each level of tree, use just one extreme range.
\ei

As shown in Table~\ref{tab:rq2}, for 13/20 experiments, untuned DART performed better than tuned RandomForests. Also, in all cases where DART performed worse,
the performance delta was very small (the largest loss was 4\% seen in the xerces' $P_{opt}$ results).

\begin{table}
    \caption {DART v.s. data-savvy learners in $\mathit{dist2heaven}$ and $P_\mathit{opt}$. Gray cells mark best performances on each project.}
    \label{tbl:RQ3_dist2heaven}
    \centering
    \footnotesize
    \resizebox{0.48\textwidth}{!}{
        \begin{tabular}{l|r|ccccccc}\hline
       \rowcolor{gray!30}  
     Goal & Data & DART & KNN & SMO  & NB  & RF & SL & DT\\ \hline
\multirow{10}{*}{\rotatebox[origin=c]{90}{\parbox[c]{3cm}{{\em dis2heaven: ({\em less} is {\em better})}}}} &       log4j & \colorbox{gray!30}{23} 	&45	&44   &50	&44	&40     &47 \\ 
      &  jedit &\colorbox{gray!30}{31}  &45	&52	&41	&39	&44	&40\\
      &   lucene & \colorbox{gray!30}{33}  	&37	&45	&44	&41	&40	&40 \\
      &  poi & \colorbox{gray!30}{35} 	& 38	& 52	& 52	& 39	& 46	& 43 \\
      &  ivy & \colorbox{gray!30}{35} 	& 37	& 46	& 36	& 39	& 37	& 40 \\
      &  velocity & \colorbox{gray!30}{37}  & 56	& 64	& 40	& 44	& 61	& 42 \\
      &  synapse & 38	&  \colorbox{gray!30}{36}	& 47	& \colorbox{gray!30}{36}	& 42 & 37	& 42 \\
      &  xalan & 39 & \colorbox{gray!30}{20}	& 35	& 45 & 	25	& 71	& 28 \\
      &  camel & 41 &	45	& 62  & 47	& \colorbox{gray!30}{35}	& 53	& 38\\ 
      &  xerces & \colorbox{gray!30}{42}	& 45	& 67	& 52	& 52	& 53	& 53 \\ \hline

       & ivy	 & \colorbox{gray!30}{28}	 & 26 & 27 & 10	& 27 & 24 & 26 \\
 \multirow{10}{*}{\rotatebox[origin=c]{90}{\parbox[c]{2cm}{$P_{\mathit{opt}}$: ({\em more} is {\em better})}}} &  jedit & \colorbox{gray!30}{39} & 3 & 17 & 6 & 10 & 4 & 24 \\
      &  synapse	& \colorbox{gray!30}{43}	& 39& 38 &	27 & 36 & 36 & 35 \\
      &  camel & 52.9 & \colorbox{gray!30}{53}	& \colorbox{gray!30}{53} & 21&	52 &	\colorbox{gray!30}{53} & 49 \\
      &  log4j &	\colorbox{gray!30}{56} &	27 &	50 &	24 &	33 &	44&	44 \\
      &  velocity &64 &	56 &	64 &	64 &	57 &	\colorbox{gray!30}{65} &	53 \\
      &  poi	 & \colorbox{gray!30}{73} &	67 &	69 & 26 & 72 & 72 &	71 \\
      &  lucene & \colorbox{gray!30}{81} &	45 &	49 &	27 &	49 &	42 &	53 \\
       & xerces	 & \colorbox{gray!30}{90} &	73 &	63 &	20 &	50 &	77 & 48 \\
      &  xalan &	99 &	99 &	98 &	24 &	93 & \colorbox{gray!30}{100} & 88 \\ \hline 
        \end{tabular}
        }

\end{table}

From these results, we assert that DART out-performs the established state-of-the-art in parameter tuning for defect prediction.
This is an interesting result since TUNER must evaluate dozens to hundreds of different models before it can select the best
settings. DART, on the other hand, just had to build one ensemble, then test one tree from that ensemble.

\subsection{RQ3: Do data-savvy learners sample results space better than a few DARTs?}

There has been much recent research in hyper-parameter tuning
of quality predictors. For example:
\bi
\item
The Fu et al. study mentioned in
{\bf RQ2} tuned control parameters of the learning algorithm.
\item
At ICSE'18, Agrawal et al.~\cite{agrawal2018better}, applied  tuning to the data pre-processor that was called before the learners
executed.
\ei
This section compares DART to the Agrawal et al. methods.
Note that the Fu et al. tuners were ``goal-savvy'',
the 
Agrawal et al. methods are ``data-savvy''.

Agrawal compared the benefits of (a)~selecting better learners versus (b)~picking any learner but also addressing class-imbalance in the
training data. 
Class-imbalance is a  major problem in quality prediction. If the target class is very rare, it can be difficult for a data mining algorithm to generate a model that can locate it.
A standard method for addressing class imbalance is the SMOTE
pre-processor~\cite{Chawla:2002}.  SMOTE randomly
deletes members of the majority class while synthesizing
artificial members of the minority class. 

SMOTE is controlled  by the parameters of Table~\ref{tbl:parameters}.
Agrawal et al. applied the same TUNER algorithm of Fu et al.
and found that the default settings of SMOTE could
be greatly improved. Agrawal et al. used the term SMOTUNED to denote their combination.

This section compares DART against SMOTUNED.
As per the methods of  Agrawal et al.~\cite{agrawal2018better},  for each project of Table\ref{tbl:data},
we randomly split the original training data into $80\%$ and $20\%$ as new training 
data and tuning data, respectively. The tuning data was used
to validate our parameter settings of SMOTE found by  TUNER. 

Similar to the RQ2 experiment, we repeated the whole process until we
either run out of tuning resources(i.e., a given tuning budget) or TUNER
could not find any better hyper-parameters. The best parameters found was tested against the testing set and these results are reported.

Since different data split might have
an impact on predictor's performance, we repeat tuning+testing process
30 times, each time with different random seed and return the median
values of 30 runs as the result of tuning SMOTE on
each project data. SMOTUNED  experiment was run twice to minimize distance2heaven and maximize popt20 metrics. 
These two experiments were run separately.

Table~\ref{tbl:RQ3_dist2heaven} shows the results.
In results that echo all the above,
usually DART performs much better than the more elaborate
approach of Agrawal et al:
\bi
\item
For the  $P_{\mathit{opt}}$ results, DART was either the best result
or no 
worse that  1\% off the best results;
\item
As to {\em dis2heaven},
DART's worst performance was for xalan, which was  was 19\% worse that best. Apart from that, DART either had the best result or was within 2\% of the best result for $\frac{8}{10}$ of the results.
\ei

\section{Threads to Validity}

As with any large scale empirical study, biases can affect the final results. Therefore, any conclusions made from this work must be considered with the following issues in mind:

 Threats to \textbf{internal validity} concern the consistency of the results obtained from the result. In our study, to investigate how DART
 performs compared with the state-of-the-art defect predictors, goal-savvy defect predictors, and data-savvy defect predictors, we has taken
 care to either clearly define our algorithms or use implementations from the public domain. All the machine learning algorithms are imported
 from Scikit-Learn, a machine learning package in Python~\cite{pedregosa2011scikit}. For example, In RQ1, DART   followed the FFTs algorithm
 defined in~\cite{martignon2003naive}. In RQ2 and RQ3, we adopt the original source code of DE-TUNER and SMOTUNED provided
 by Fu et al~\cite{fu2016tuning} and Agrawal et. al.~\cite{agrawal2018better}, which reduce the bias introduced by implementing the rigs by
 ourselves. All the data used in this work is widely used open source Java system
 data  in defect prediction field and it is also available in the SEACRAFT repository~(http://tiny.cc/seacraft).

 Threats to \textbf{external validity} represent if the results are of relevance for other cases, or the ability to generalize the observations in a study.
 In this study, we proposed that using DART as a scout to explore the results space could build better defect predictors in terms of $dis2heaven$ and $P_{opt}$ measures. Nonetheless, we do not claim that our findings can be generalized to all software quality predictors tasks. However, those other software quality predictors tasks often apply machine learning algorithms, like SVM and Random Forests, or other data pre-processing techniques to build predictive models.
 Most of those models are also exploring the results space and find the best models. Therefore, it is quite possible that FFTs method of this paper  would be widely applicable, elsewhere.

\section{Conclusions}\label{discussion}

The thesis of this paper is that we have been treating
uncertainty incorrectly. Instead of view uncertainty 
as a problems to be solved, we instead view it as a resource
that simplifies software defect prediction.

For example, 
Deb's principle of $\epsilon$-dominance
  states that if there exists some
$\epsilon$
value below which
it is useless or impossible to distinguish results, then
it is superfluous to explore anything less than
$\epsilon$. For large $\epsilon$ problems, the
 results space effectively contains just a few regions.
 
 As shown here, there are several important benefits
 if we we design a learner especially for
 such large $\epsilon$ problems. Firstly,
 the resulting learner is very simple to implement.
 Secondly, this learner can sample the results space
 very effectively. Thirdly, that very simple learner
 can out-perform far more elaborate systems such as
  three state-of-the-art defect prediction systems:
\begin{enumerate}
\item
The   algorithms surveyed at a recent  ICSE'15 paper~\cite{ghotra2015revisiting};
\item
A hyper-parameter optimization method  proposed in 2016 in the IST journal~\cite{fu2016tuning}; 
\item
A search-based data pre-processing method presented at ICSE'18~\cite{agrawal2018better}. 
\end{enumerate}

\begin{figure}[!t]
\centerline{\includegraphics[width=\linewidth]{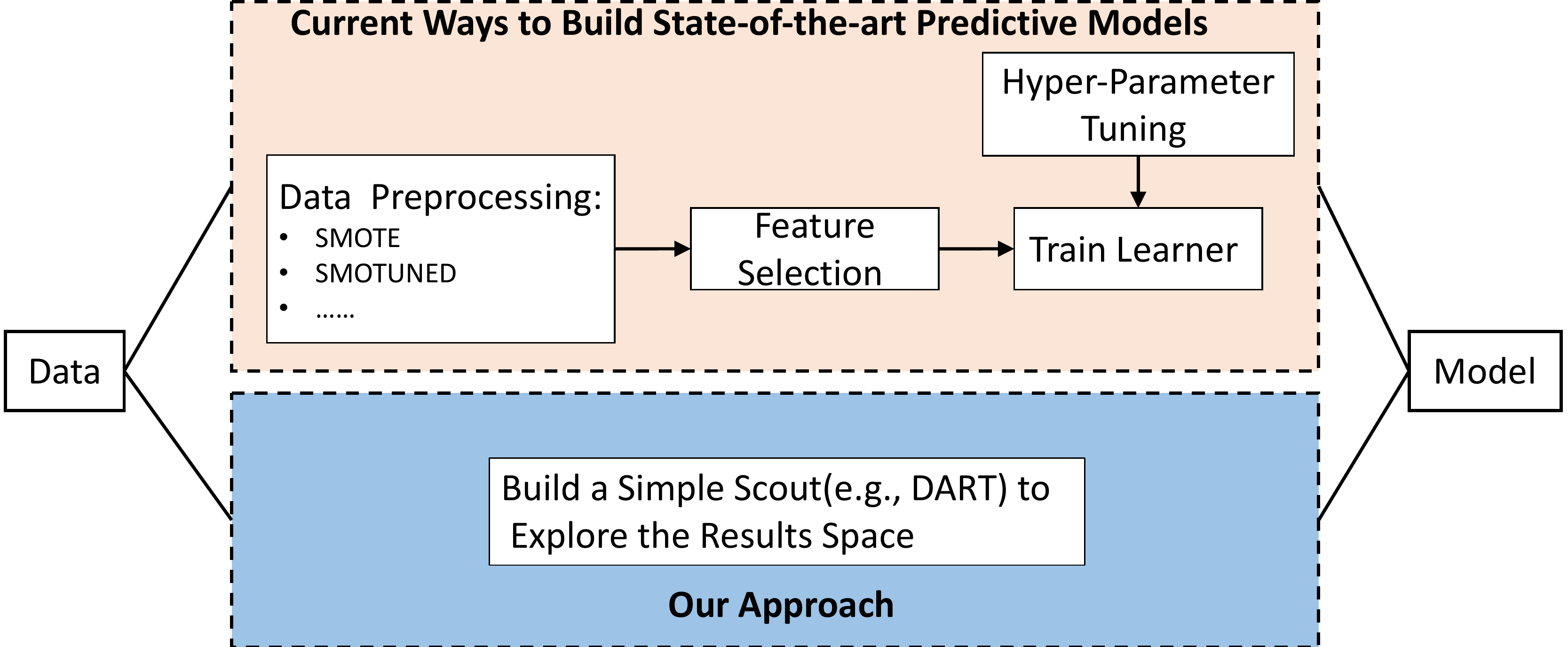}}
\caption{Different ways to 
reason about software quality prediction.}\label{fig:ways}
\end{figure}
  
We believe that our results call for a new approach to software quality prediction.
The standard approach to this problem, as shown by the top pink section of \fig{ways} is to reason
{\em forwards} from  domain data, towards a model. In that approach, analysts must make many decisions
about  data pre-processing, feature selection, and tuning parameters for a learner. This is a very large number
of decisions:
\begin{itemize}
\item
Data can be processed by SMOTE~\cite{Chawla:2002}, SMOTUNED~\cite{agrawal2018better}, or any number of other methods
including normalization, discretization, outlier removal, etc~\cite{Witten:2005};
\item 
Feature selection can explore $2^N$ subsets of $N$ features;
\item
Hyper-parameter optimization  explores the space of control parameters within data mining algorithms.
As shown in Table~\ref{tbl:parameters}, those parameters can be continuous which means the space of 
parameters is theoretically infinite.
\end{itemize}
Perhaps a much simpler approach is  
the {\em backwards reasoning} shown in the blue bottom
region of \fig{ways}. In this approach, analysts
do some initial data mining, perhaps at random,
then reflect on what has been learned from
those initial probes of the result space.
Based on those results, analysts then design a software
quality predictor that better understands the results space. 

The DART system discussed in this paper is an example of such
backwards reasoning. We hope the success of this system
inspires other researchers to explore
large scale simplifications of other SE problem domains.


\bibliographystyle{ACM-Reference-Format}
\balance

\begin{thebibliography}{00}


\ifx \showCODEN    \undefined \def \showCODEN     #1{\unskip}     \fi
\ifx \showDOI      \undefined \def \showDOI       #1{{\tt DOI:}\penalty0{#1}\ }
  \fi
\ifx \showISBNx    \undefined \def \showISBNx     #1{\unskip}     \fi
\ifx \showISBNxiii \undefined \def \showISBNxiii  #1{\unskip}     \fi
\ifx \showISSN     \undefined \def \showISSN      #1{\unskip}     \fi
\ifx \showLCCN     \undefined \def \showLCCN      #1{\unskip}     \fi
\ifx \shownote     \undefined \def \shownote      #1{#1}          \fi
\ifx \showarticletitle \undefined \def \showarticletitle #1{#1}   \fi
\ifx \showURL      \undefined \def \showURL       #1{#1}          \fi
\providecommand\bibfield[2]{#2}
\providecommand\bibinfo[2]{#2}
\providecommand\natexlab[1]{#1}
\providecommand\showeprint[2][]{arXiv:#2}

\bibitem[\protect\citeauthoryear{??}{che}{2018}]%
        {chen2018multi}
 \bibinfo{year}{2018}\natexlab{}.
\newblock \showarticletitle{MULTI: Multi-objective effort-aware just-in-time
  software defect prediction}.
\newblock \bibinfo{journal}{{\em Information and Software Technology\/}}
  \bibinfo{volume}{93} (\bibinfo{year}{2018}), \bibinfo{pages}{1--13}.
\newblock


\bibitem[\protect\citeauthoryear{Agrawal, Fu, and Menzies}{Agrawal
  et~al\mbox{.}}{2018}]%
        {Agrawal18stab}
\bibfield{author}{\bibinfo{person}{Amritanshu Agrawal}, \bibinfo{person}{Wei
  Fu}, {and} \bibinfo{person}{Tim Menzies}.} \bibinfo{year}{2018}\natexlab{}.
\newblock \showarticletitle{What is Wrong with Topic Modeling? (and How to Fix
  it Using Search-based SE)}.
\newblock  (\bibinfo{date}{02} \bibinfo{year}{2018}).
\newblock


\bibitem[\protect\citeauthoryear{Agrawal and Menzies}{Agrawal and
  Menzies}{2018}]%
        {agrawal2018better}
\bibfield{author}{\bibinfo{person}{Amritanshu Agrawal} {and}
  \bibinfo{person}{Tim Menzies}.} \bibinfo{year}{2018}\natexlab{}.
\newblock \showarticletitle{Is "Better Data" Better than "Better Data Miners"?
  (Benefits of Tuning SMOTE for Defect Prediction)}.
\newblock \bibinfo{journal}{{\em ICSE\/}} (\bibinfo{year}{2018}).
\newblock


\bibitem[\protect\citeauthoryear{Barr, Harman, McMinn, Shahbaz, and Yoo}{Barr
  et~al\mbox{.}}{2015}]%
        {barr2015oracle}
\bibfield{author}{\bibinfo{person}{Earl~T Barr}, \bibinfo{person}{Mark Harman},
  \bibinfo{person}{Phil McMinn}, \bibinfo{person}{Muzammil Shahbaz}, {and}
  \bibinfo{person}{Shin Yoo}.} \bibinfo{year}{2015}\natexlab{}.
\newblock \showarticletitle{The oracle problem in software testing: A survey}.
\newblock \bibinfo{journal}{{\em IEEE transactions on software engineering\/}}
  \bibinfo{volume}{41}, \bibinfo{number}{5} (\bibinfo{year}{2015}),
  \bibinfo{pages}{507--525}.
\newblock


\bibitem[\protect\citeauthoryear{Calero and Piattini}{Calero and
  Piattini}{2015}]%
        {Calero:2015}
\bibfield{author}{\bibinfo{person}{Coral Calero} {and} \bibinfo{person}{Mario
  Piattini}.} \bibinfo{year}{2015}\natexlab{}.
\newblock \bibinfo{booktitle}{{\em Green in Software Engineering}}.
\newblock \bibinfo{publisher}{Springer Publishing Company, Incorporated}.
\newblock
\showISBNx{3319085808, 9783319085807}


\bibitem[\protect\citeauthoryear{Chawla, Bowyer, Hall, and Kegelmeyer}{Chawla
  et~al\mbox{.}}{2002}]%
        {Chawla:2002}
\bibfield{author}{\bibinfo{person}{Nitesh~V. Chawla}, \bibinfo{person}{Kevin~W.
  Bowyer}, \bibinfo{person}{Lawrence~O. Hall}, {and} \bibinfo{person}{W.~Philip
  Kegelmeyer}.} \bibinfo{year}{2002}\natexlab{}.
\newblock \showarticletitle{SMOTE: Synthetic Minority Over-sampling Technique}.
\newblock \bibinfo{journal}{{\em J. Artif. Int. Res.\/}} \bibinfo{volume}{16},
  \bibinfo{number}{1} (\bibinfo{date}{June} \bibinfo{year}{2002}),
  \bibinfo{pages}{321--357}.
\newblock
\showISSN{1076-9757}
\showURL{%
\url{http://dl.acm.org/citation.cfm?id=1622407.1622416}}


\bibitem[\protect\citeauthoryear{Chen, Menzies, Port, and Boehm}{Chen
  et~al\mbox{.}}{2005}]%
        {chen2005feature}
\bibfield{author}{\bibinfo{person}{Zhihao Chen}, \bibinfo{person}{Tim Menzies},
  \bibinfo{person}{Dan Port}, {and} \bibinfo{person}{Barry Boehm}.}
  \bibinfo{year}{2005}\natexlab{}.
\newblock \showarticletitle{Feature subset selection can improve software cost
  estimation accuracy}. In \bibinfo{booktitle}{{\em ACM SIGSOFT Software
  Engineering Notes}}, Vol.~\bibinfo{volume}{30}. ACM, \bibinfo{pages}{1--6}.
\newblock


\bibitem[\protect\citeauthoryear{Chidamber and Kemerer}{Chidamber and
  Kemerer}{1994}]%
        {chidamber1994metrics}
\bibfield{author}{\bibinfo{person}{Shyam~R Chidamber} {and}
  \bibinfo{person}{Chris~F Kemerer}.} \bibinfo{year}{1994}\natexlab{}.
\newblock \showarticletitle{A metrics suite for object oriented design}.
\newblock \bibinfo{journal}{{\em IEEE Transactions on software engineering\/}}
  \bibinfo{volume}{20}, \bibinfo{number}{6} (\bibinfo{year}{1994}),
  \bibinfo{pages}{476--493}.
\newblock


\bibitem[\protect\citeauthoryear{Deb, Mohan, and Mishra}{Deb
  et~al\mbox{.}}{2005}]%
        {deb2005evaluating}
\bibfield{author}{\bibinfo{person}{Kalyanmoy Deb}, \bibinfo{person}{Manikanth
  Mohan}, {and} \bibinfo{person}{Shikhar Mishra}.}
  \bibinfo{year}{2005}\natexlab{}.
\newblock \showarticletitle{Evaluating the $\varepsilon$-domination based
  multi-objective evolutionary algorithm for a quick computation of
  Pareto-optimal solutions}.
\newblock \bibinfo{journal}{{\em Evolutionary computation\/}}
  \bibinfo{volume}{13}, \bibinfo{number}{4} (\bibinfo{year}{2005}),
  \bibinfo{pages}{501--525}.
\newblock


\bibitem[\protect\citeauthoryear{Fayyad and Irani}{Fayyad and Irani}{1993}]%
        {fayyad1993multi}
\bibfield{author}{\bibinfo{person}{Usama Fayyad} {and} \bibinfo{person}{Keki
  Irani}.} \bibinfo{year}{1993}\natexlab{}.
\newblock \showarticletitle{Multi-interval discretization of continuous-valued
  attributes for classification learning}.
\newblock  (\bibinfo{year}{1993}).
\newblock


\bibitem[\protect\citeauthoryear{Fenton and Neil}{Fenton and Neil}{2012}]%
        {fenton2012risk}
\bibfield{author}{\bibinfo{person}{Norman Fenton} {and} \bibinfo{person}{Martin
  Neil}.} \bibinfo{year}{2012}\natexlab{}.
\newblock \bibinfo{booktitle}{{\em Risk assessment and decision analysis with
  Bayesian networks}}.
\newblock \bibinfo{publisher}{Crc Press}.
\newblock


\bibitem[\protect\citeauthoryear{Fu and Menzies}{Fu and Menzies}{2017a}]%
        {fu2017easy}
\bibfield{author}{\bibinfo{person}{Wei Fu} {and} \bibinfo{person}{Tim
  Menzies}.} \bibinfo{year}{2017}\natexlab{a}.
\newblock \showarticletitle{Easy over Hard: A Case Study on Deep Learning}. In
  \bibinfo{booktitle}{{\em Proceedings of the 2017 11th Joint Meeting on
  Foundations of Software Engineering}} {\em (\bibinfo{series}{ESEC/FSE
  2017})}. \bibinfo{publisher}{ACM}, \bibinfo{pages}{49--60}.
\newblock


\bibitem[\protect\citeauthoryear{Fu and Menzies}{Fu and Menzies}{2017b}]%
        {fu2017unsupervised}
\bibfield{author}{\bibinfo{person}{Wei Fu} {and} \bibinfo{person}{Tim
  Menzies}.} \bibinfo{year}{2017}\natexlab{b}.
\newblock \showarticletitle{Revisiting Unsupervised Learning for Defect
  Prediction}. In \bibinfo{booktitle}{{\em Proceedings of the 2017 11th Joint
  Meeting on Foundations of Software Engineering}} {\em
  (\bibinfo{series}{ESEC/FSE 2017})}. \bibinfo{publisher}{ACM},
  \bibinfo{pages}{72--83}.
\newblock


\bibitem[\protect\citeauthoryear{Fu, Menzies, and Shen}{Fu
  et~al\mbox{.}}{2016a}]%
        {fu2016tuning}
\bibfield{author}{\bibinfo{person}{Wei Fu}, \bibinfo{person}{Tim Menzies},
  {and} \bibinfo{person}{Xipeng Shen}.} \bibinfo{year}{2016}\natexlab{a}.
\newblock \showarticletitle{Tuning for software analytics: Is it really
  necessary?}
\newblock \bibinfo{journal}{{\em Information and Software Technology\/}}
  \bibinfo{volume}{76} (\bibinfo{year}{2016}), \bibinfo{pages}{135--146}.
\newblock


\bibitem[\protect\citeauthoryear{Fu, Nair, and Menzies}{Fu
  et~al\mbox{.}}{2016b}]%
        {fu2016differential}
\bibfield{author}{\bibinfo{person}{Wei Fu}, \bibinfo{person}{Vivek Nair}, {and}
  \bibinfo{person}{Tim Menzies}.} \bibinfo{year}{2016}\natexlab{b}.
\newblock \showarticletitle{Why is differential evolution better than grid
  search for tuning defect predictors?}
\newblock \bibinfo{journal}{{\em arXiv preprint arXiv:1609.02613\/}}
  (\bibinfo{year}{2016}).
\newblock


\bibitem[\protect\citeauthoryear{Ghotra, McIntosh, and Hassan}{Ghotra
  et~al\mbox{.}}{2015}]%
        {ghotra2015revisiting}
\bibfield{author}{\bibinfo{person}{Baljinder Ghotra}, \bibinfo{person}{Shane
  McIntosh}, {and} \bibinfo{person}{Ahmed~E Hassan}.}
  \bibinfo{year}{2015}\natexlab{}.
\newblock \showarticletitle{Revisiting the impact of classification techniques
  on the performance of defect prediction models}. IEEE Press,
  \bibinfo{pages}{789--800}.
\newblock


\bibitem[\protect\citeauthoryear{Hall, Beecham, Bowes, Gray, and Counsell}{Hall
  et~al\mbox{.}}{2012}]%
        {hall2012systematic}
\bibfield{author}{\bibinfo{person}{Tracy Hall}, \bibinfo{person}{Sarah
  Beecham}, \bibinfo{person}{David Bowes}, \bibinfo{person}{David Gray}, {and}
  \bibinfo{person}{Steve Counsell}.} \bibinfo{year}{2012}\natexlab{}.
\newblock \showarticletitle{A systematic literature review on fault prediction
  performance in software engineering}.
\newblock \bibinfo{journal}{{\em IEEE Transactions on Software Engineering\/}}
  \bibinfo{volume}{38}, \bibinfo{number}{6} (\bibinfo{year}{2012}),
  \bibinfo{pages}{1276--1304}.
\newblock


\bibitem[\protect\citeauthoryear{Hamill and Goseva-Popstojanova}{Hamill and
  Goseva-Popstojanova}{2009}]%
        {hamill2009common}
\bibfield{author}{\bibinfo{person}{Maggie Hamill} {and}
  \bibinfo{person}{Katerina Goseva-Popstojanova}.}
  \bibinfo{year}{2009}\natexlab{}.
\newblock \showarticletitle{Common trends in software fault and failure data}.
\newblock \bibinfo{journal}{{\em IEEE Transactions on Software Engineering\/}}
  \bibinfo{volume}{35}, \bibinfo{number}{4} (\bibinfo{year}{2009}),
  \bibinfo{pages}{484--496}.
\newblock


\bibitem[\protect\citeauthoryear{Hosseini, Turhan, and Gunarathna}{Hosseini
  et~al\mbox{.}}{2017}]%
        {hosseini2017systematic}
\bibfield{author}{\bibinfo{person}{Seyedrebvar Hosseini},
  \bibinfo{person}{Burak Turhan}, {and} \bibinfo{person}{Dimuthu Gunarathna}.}
  \bibinfo{year}{2017}\natexlab{}.
\newblock \showarticletitle{A Systematic Literature Review and Meta-analysis on
  Cross Project Defect Prediction}.
\newblock \bibinfo{journal}{{\em IEEE Transactions on Software Engineering\/}}
  (\bibinfo{year}{2017}).
\newblock


\bibitem[\protect\citeauthoryear{Kafura and Reddy}{Kafura and Reddy}{1987}]%
        {kafura1987use}
\bibfield{author}{\bibinfo{person}{Dennis Kafura} {and}
  \bibinfo{person}{Geereddy~R. Reddy}.} \bibinfo{year}{1987}\natexlab{}.
\newblock \showarticletitle{The use of software complexity metrics in software
  maintenance}.
\newblock \bibinfo{journal}{{\em IEEE Transactions on Software Engineering\/}}
  \bibinfo{number}{3} (\bibinfo{year}{1987}), \bibinfo{pages}{335--343}.
\newblock


\bibitem[\protect\citeauthoryear{Kamei, Shihab, Adams, Hassan, Mockus, Sinha,
  and Ubayashi}{Kamei et~al\mbox{.}}{2013}]%
        {kamei2013large}
\bibfield{author}{\bibinfo{person}{Yasutaka Kamei}, \bibinfo{person}{Emad
  Shihab}, \bibinfo{person}{Bram Adams}, \bibinfo{person}{Ahmed~E Hassan},
  \bibinfo{person}{Audris Mockus}, \bibinfo{person}{Anand Sinha}, {and}
  \bibinfo{person}{Naoyasu Ubayashi}.} \bibinfo{year}{2013}\natexlab{}.
\newblock \showarticletitle{A large-scale empirical study of just-in-time
  quality assurance}.
\newblock \bibinfo{journal}{{\em IEEE Transactions on Software Engineering\/}}
  \bibinfo{volume}{39}, \bibinfo{number}{6} (\bibinfo{year}{2013}),
  \bibinfo{pages}{757--773}.
\newblock


\bibitem[\protect\citeauthoryear{Khoshgoftaar and Allen}{Khoshgoftaar and
  Allen}{2001}]%
        {khoshgoftaar2001modeling}
\bibfield{author}{\bibinfo{person}{Taghi~M Khoshgoftaar} {and}
  \bibinfo{person}{Edward~B Allen}.} \bibinfo{year}{2001}\natexlab{}.
\newblock \showarticletitle{Modeling software quality with}.
\newblock \bibinfo{journal}{{\em Recent Advances in Reliability and Quality
  Engineering\/}}  \bibinfo{volume}{2} (\bibinfo{year}{2001}),
  \bibinfo{pages}{247}.
\newblock


\bibitem[\protect\citeauthoryear{Khoshgoftaar and Seliya}{Khoshgoftaar and
  Seliya}{2003}]%
        {khoshgoftaar2003software}
\bibfield{author}{\bibinfo{person}{Taghi~M Khoshgoftaar} {and}
  \bibinfo{person}{Naeem Seliya}.} \bibinfo{year}{2003}\natexlab{}.
\newblock \showarticletitle{Software quality classification modeling using the
  SPRINT decision tree algorithm}.
\newblock \bibinfo{journal}{{\em International Journal on Artificial
  Intelligence Tools\/}} \bibinfo{volume}{12}, \bibinfo{number}{03}
  (\bibinfo{year}{2003}), \bibinfo{pages}{207--225}.
\newblock


\bibitem[\protect\citeauthoryear{Khoshgoftaar, Yuan, and Allen}{Khoshgoftaar
  et~al\mbox{.}}{2000}]%
        {khoshgoftaar2000balancing}
\bibfield{author}{\bibinfo{person}{Taghi~M Khoshgoftaar},
  \bibinfo{person}{Xiaojing Yuan}, {and} \bibinfo{person}{Edward~B Allen}.}
  \bibinfo{year}{2000}\natexlab{}.
\newblock \showarticletitle{Balancing misclassification rates in
  classification-tree models of software quality}.
\newblock \bibinfo{journal}{{\em Empirical Software Engineering\/}}
  \bibinfo{volume}{5}, \bibinfo{number}{4} (\bibinfo{year}{2000}),
  \bibinfo{pages}{313--330}.
\newblock


\bibitem[\protect\citeauthoryear{Kim, Nam, Yeon, Choi, and Kim}{Kim
  et~al\mbox{.}}{2015}]%
        {kim2015remi}
\bibfield{author}{\bibinfo{person}{Mijung Kim}, \bibinfo{person}{Jaechang Nam},
  \bibinfo{person}{Jaehyuk Yeon}, \bibinfo{person}{Soonhwang Choi}, {and}
  \bibinfo{person}{Sunghun Kim}.} \bibinfo{year}{2015}\natexlab{}.
\newblock \showarticletitle{REMI: defect prediction for efficient API testing}.
  In \bibinfo{booktitle}{{\em Proceedings of the 2015 10th Joint Meeting on
  Foundations of Software Engineering}}. ACM, \bibinfo{pages}{990--993}.
\newblock


\bibitem[\protect\citeauthoryear{Kim, Whitehead~Jr, and Zhang}{Kim
  et~al\mbox{.}}{2008}]%
        {kim2008classifying}
\bibfield{author}{\bibinfo{person}{Sunghun Kim}, \bibinfo{person}{E~James
  Whitehead~Jr}, {and} \bibinfo{person}{Yi Zhang}.}
  \bibinfo{year}{2008}\natexlab{}.
\newblock \showarticletitle{Classifying software changes: Clean or buggy?}
\newblock \bibinfo{journal}{{\em IEEE Transactions on Software Engineering\/}}
  \bibinfo{volume}{34}, \bibinfo{number}{2} (\bibinfo{year}{2008}),
  \bibinfo{pages}{181--196}.
\newblock


\bibitem[\protect\citeauthoryear{Kocaguneli, Gay, Menzies, Yang, and
  Keung}{Kocaguneli et~al\mbox{.}}{2010}]%
        {kocaguneli2010use}
\bibfield{author}{\bibinfo{person}{Ekrem Kocaguneli}, \bibinfo{person}{Gregory
  Gay}, \bibinfo{person}{Tim Menzies}, \bibinfo{person}{Ye Yang}, {and}
  \bibinfo{person}{Jacky~W Keung}.} \bibinfo{year}{2010}\natexlab{}.
\newblock \showarticletitle{When to use data from other projects for effort
  estimation}. In \bibinfo{booktitle}{{\em Proceedings of the IEEE/ACM
  international conference on Automated software engineering}}. ACM,
  \bibinfo{pages}{321--324}.
\newblock


\bibitem[\protect\citeauthoryear{Koru, Zhang, El~Emam, and Liu}{Koru
  et~al\mbox{.}}{2009}]%
        {koru2009investigation}
\bibfield{author}{\bibinfo{person}{A~G{\"u}ne{\c{s}} Koru},
  \bibinfo{person}{Dongsong Zhang}, \bibinfo{person}{Khaled El~Emam}, {and}
  \bibinfo{person}{Hongfang Liu}.} \bibinfo{year}{2009}\natexlab{}.
\newblock \showarticletitle{An investigation into the functional form of the
  size-defect relationship for software modules}.
\newblock \bibinfo{journal}{{\em IEEE Transactions on Software Engineering\/}}
  \bibinfo{volume}{35}, \bibinfo{number}{2} (\bibinfo{year}{2009}),
  \bibinfo{pages}{293--304}.
\newblock


\bibitem[\protect\citeauthoryear{Lessmann, Baesens, Mues, and Pietsch}{Lessmann
  et~al\mbox{.}}{2008a}]%
        {Lessmann08}
\bibfield{author}{\bibinfo{person}{S. Lessmann}, \bibinfo{person}{B. Baesens},
  \bibinfo{person}{C. Mues}, {and} \bibinfo{person}{S. Pietsch}.}
  \bibinfo{year}{2008}\natexlab{a}.
\newblock \showarticletitle{Benchmarking Classification Models for Software
  Defect Prediction: A Proposed Framework and Novel Findings}.
\newblock \bibinfo{journal}{{\em IEEE Transactions on Software Engineering\/}}
  \bibinfo{volume}{34}, \bibinfo{number}{4} (\bibinfo{date}{July}
  \bibinfo{year}{2008}), \bibinfo{pages}{485--496}.
\newblock
\showISSN{0098-5589}
\showDOI{%
\url{http://dx.doi.org/10.1109/TSE.2008.35}}


\bibitem[\protect\citeauthoryear{Lessmann, Baesens, Mues, and Pietsch}{Lessmann
  et~al\mbox{.}}{2008b}]%
        {lessmann2008benchmarking}
\bibfield{author}{\bibinfo{person}{Stefan Lessmann}, \bibinfo{person}{Bart
  Baesens}, \bibinfo{person}{Christophe Mues}, {and} \bibinfo{person}{Swantje
  Pietsch}.} \bibinfo{year}{2008}\natexlab{b}.
\newblock \showarticletitle{Benchmarking classification models for software
  defect prediction: A proposed framework and novel findings}.
\newblock \bibinfo{journal}{{\em IEEE Transactions on Software Engineering\/}}
  \bibinfo{volume}{34}, \bibinfo{number}{4} (\bibinfo{year}{2008}),
  \bibinfo{pages}{485--496}.
\newblock


\bibitem[\protect\citeauthoryear{Martignon, Vitouch, Takezawa, and
  Forster}{Martignon et~al\mbox{.}}{2003}]%
        {martignon2003naive}
\bibfield{author}{\bibinfo{person}{Laura Martignon}, \bibinfo{person}{Oliver
  Vitouch}, \bibinfo{person}{Masanori Takezawa}, {and}
  \bibinfo{person}{Malcolm~R Forster}.} \bibinfo{year}{2003}\natexlab{}.
\newblock \showarticletitle{Naive and yet enlightened: From natural frequencies
  to fast and frugal decision trees}.
\newblock \bibinfo{journal}{{\em Thinking: Psychological perspectives on
  reasoning, judgment and decision making\/}} (\bibinfo{year}{2003}),
  \bibinfo{pages}{189--211}.
\newblock


\bibitem[\protect\citeauthoryear{McCabe}{McCabe}{1976}]%
        {mccabe1976complexity}
\bibfield{author}{\bibinfo{person}{Thomas~J McCabe}.}
  \bibinfo{year}{1976}\natexlab{}.
\newblock \showarticletitle{A complexity measure}.
\newblock \bibinfo{journal}{{\em IEEE Transactions on software Engineering\/}}
  \bibinfo{number}{4} (\bibinfo{year}{1976}), \bibinfo{pages}{308--320}.
\newblock


\bibitem[\protect\citeauthoryear{Menzies, Dekhtyar, Distefano, and
  Greenwald}{Menzies et~al\mbox{.}}{2007a}]%
        {Menzies:2007prec}
\bibfield{author}{\bibinfo{person}{Tim Menzies}, \bibinfo{person}{Alex
  Dekhtyar}, \bibinfo{person}{Justin Distefano}, {and} \bibinfo{person}{Jeremy
  Greenwald}.} \bibinfo{year}{2007}\natexlab{a}.
\newblock \showarticletitle{Problems with Precision: A Response to "Comments on
  'Data Mining Static Code Attributes to Learn Defect Predictors'"}.
\newblock \bibinfo{journal}{{\em IEEE Trans. Softw. Eng.\/}}
  \bibinfo{volume}{33}, \bibinfo{number}{9} (\bibinfo{date}{Sept.}
  \bibinfo{year}{2007}), \bibinfo{pages}{637--640}.
\newblock
\showISSN{0098-5589}
\showDOI{%
\url{http://dx.doi.org/10.1109/TSE.2007.70721}}


\bibitem[\protect\citeauthoryear{Menzies, Greenwald, and Frank}{Menzies
  et~al\mbox{.}}{2007b}]%
        {menzies2007data}
\bibfield{author}{\bibinfo{person}{Tim Menzies}, \bibinfo{person}{Jeremy
  Greenwald}, {and} \bibinfo{person}{Art Frank}.}
  \bibinfo{year}{2007}\natexlab{b}.
\newblock \showarticletitle{Data mining static code attributes to learn defect
  predictors}.
\newblock \bibinfo{journal}{{\em IEEE Transactions on Software Engineering\/}}
  \bibinfo{volume}{33}, \bibinfo{number}{1} (\bibinfo{year}{2007}).
\newblock


\bibitem[\protect\citeauthoryear{Menzies and Shepperd}{Menzies and
  Shepperd}{2012}]%
        {menzies2012special}
\bibfield{author}{\bibinfo{person}{Tim Menzies} {and} \bibinfo{person}{Martin
  Shepperd}.} \bibinfo{year}{2012}\natexlab{}.
\newblock \bibinfo{title}{Special issue on repeatable results in software
  engineering prediction}.
\newblock   (\bibinfo{year}{2012}).
\newblock


\bibitem[\protect\citeauthoryear{Misirli, Bener, and Kale}{Misirli
  et~al\mbox{.}}{2011}]%
        {misirli2011ai}
\bibfield{author}{\bibinfo{person}{Ayse~Tosun Misirli}, \bibinfo{person}{Ayse
  Bener}, {and} \bibinfo{person}{Resat Kale}.} \bibinfo{year}{2011}\natexlab{}.
\newblock \showarticletitle{Ai-based software defect predictors: Applications
  and benefits in a case study}.
\newblock \bibinfo{journal}{{\em AI Magazine\/}} \bibinfo{volume}{32},
  \bibinfo{number}{2} (\bibinfo{year}{2011}), \bibinfo{pages}{57--68}.
\newblock


\bibitem[\protect\citeauthoryear{Mockus and Weiss}{Mockus and Weiss}{2000}]%
        {mockus2000predicting}
\bibfield{author}{\bibinfo{person}{Audris Mockus} {and}
  \bibinfo{person}{David~M Weiss}.} \bibinfo{year}{2000}\natexlab{}.
\newblock \showarticletitle{Predicting risk of software changes}.
\newblock \bibinfo{journal}{{\em Bell Labs Technical Journal\/}}
  \bibinfo{volume}{5}, \bibinfo{number}{2} (\bibinfo{year}{2000}),
  \bibinfo{pages}{169--180}.
\newblock


\bibitem[\protect\citeauthoryear{Monden, Hayashi, Shinoda, Shirai, Yoshida,
  Barker, and Matsumoto}{Monden et~al\mbox{.}}{2013}]%
        {monden2013assessing}
\bibfield{author}{\bibinfo{person}{Akito Monden}, \bibinfo{person}{Takuma
  Hayashi}, \bibinfo{person}{Shoji Shinoda}, \bibinfo{person}{Kumiko Shirai},
  \bibinfo{person}{Junichi Yoshida}, \bibinfo{person}{Mike Barker}, {and}
  \bibinfo{person}{Kenichi Matsumoto}.} \bibinfo{year}{2013}\natexlab{}.
\newblock \showarticletitle{Assessing the cost effectiveness of fault
  prediction in acceptance testing}.
\newblock \bibinfo{journal}{{\em IEEE Transactions on Software Engineering\/}}
  \bibinfo{volume}{39}, \bibinfo{number}{10} (\bibinfo{year}{2013}),
  \bibinfo{pages}{1345--1357}.
\newblock


\bibitem[\protect\citeauthoryear{Myers, Sandler, and Badgett}{Myers
  et~al\mbox{.}}{2011}]%
        {myers2011art}
\bibfield{author}{\bibinfo{person}{Glenford~J Myers}, \bibinfo{person}{Corey
  Sandler}, {and} \bibinfo{person}{Tom Badgett}.}
  \bibinfo{year}{2011}\natexlab{}.
\newblock \bibinfo{booktitle}{{\em The art of software testing}}.
\newblock \bibinfo{publisher}{John Wiley \& Sons}.
\newblock


\bibitem[\protect\citeauthoryear{Orso and Rothermel}{Orso and
  Rothermel}{2014}]%
        {orso2014software}
\bibfield{author}{\bibinfo{person}{Alessandro Orso} {and}
  \bibinfo{person}{Gregg Rothermel}.} \bibinfo{year}{2014}\natexlab{}.
\newblock \showarticletitle{Software testing: a research travelogue
  (2000--2014)}. In \bibinfo{booktitle}{{\em Proceedings of the on Future of
  Software Engineering}}. ACM, \bibinfo{pages}{117--132}.
\newblock


\bibitem[\protect\citeauthoryear{Ostrand, Weyuker, and Bell}{Ostrand
  et~al\mbox{.}}{2004}]%
        {ostrand2004bugs}
\bibfield{author}{\bibinfo{person}{Thomas~J Ostrand}, \bibinfo{person}{Elaine~J
  Weyuker}, {and} \bibinfo{person}{Robert~M Bell}.}
  \bibinfo{year}{2004}\natexlab{}.
\newblock \showarticletitle{Where the bugs are}. In \bibinfo{booktitle}{{\em
  ACM SIGSOFT Software Engineering Notes}}, Vol.~\bibinfo{volume}{29}. ACM,
  \bibinfo{pages}{86--96}.
\newblock


\bibitem[\protect\citeauthoryear{Pedregosa, Varoquaux, Gramfort, Michel,
  Thirion, Grisel, Blondel, Prettenhofer, Weiss, Dubourg,
  et~al\mbox{.}}{Pedregosa et~al\mbox{.}}{2011}]%
        {pedregosa2011scikit}
\bibfield{author}{\bibinfo{person}{Fabian Pedregosa}, \bibinfo{person}{Ga{\"e}l
  Varoquaux}, \bibinfo{person}{Alexandre Gramfort}, \bibinfo{person}{Vincent
  Michel}, \bibinfo{person}{Bertrand Thirion}, \bibinfo{person}{Olivier
  Grisel}, \bibinfo{person}{Mathieu Blondel}, \bibinfo{person}{Peter
  Prettenhofer}, \bibinfo{person}{Ron Weiss}, \bibinfo{person}{Vincent
  Dubourg}, {and} \bibinfo{person}{others}.} \bibinfo{year}{2011}\natexlab{}.
\newblock \showarticletitle{Scikit-learn: Machine learning in Python}.
\newblock \bibinfo{journal}{{\em Journal of machine learning research\/}}
  \bibinfo{volume}{12}, \bibinfo{number}{Oct} (\bibinfo{year}{2011}),
  \bibinfo{pages}{2825--2830}.
\newblock


\bibitem[\protect\citeauthoryear{Phillips, Neth, Woike, and
  Gaissmaier}{Phillips et~al\mbox{.}}{2017}]%
        {phillips2017fftrees}
\bibfield{author}{\bibinfo{person}{Nathaniel~D Phillips},
  \bibinfo{person}{Hansj{\"o}rg Neth}, \bibinfo{person}{Jan~K Woike}, {and}
  \bibinfo{person}{Wolfgang Gaissmaier}.} \bibinfo{year}{2017}\natexlab{}.
\newblock \showarticletitle{FFTrees: A toolbox to create, visualize, and
  evaluate fast-and-frugal decision trees}.
\newblock \bibinfo{journal}{{\em Judgment and Decision Making\/}}
  \bibinfo{volume}{12}, \bibinfo{number}{4} (\bibinfo{year}{2017}),
  \bibinfo{pages}{344}.
\newblock


\bibitem[\protect\citeauthoryear{Rahman and Devanbu}{Rahman and
  Devanbu}{2013}]%
        {rahman2013and}
\bibfield{author}{\bibinfo{person}{Foyzur Rahman} {and}
  \bibinfo{person}{Premkumar Devanbu}.} \bibinfo{year}{2013}\natexlab{}.
\newblock \showarticletitle{How, and why, process metrics are better}. In
  \bibinfo{booktitle}{{\em Software Engineering (ICSE), 2013 35th International
  Conference on}}. IEEE, \bibinfo{pages}{432--441}.
\newblock


\bibitem[\protect\citeauthoryear{Rahman, Khatri, Barr, and Devanbu}{Rahman
  et~al\mbox{.}}{2014}]%
        {rahman2014comparing}
\bibfield{author}{\bibinfo{person}{Foyzur Rahman}, \bibinfo{person}{Sameer
  Khatri}, \bibinfo{person}{Earl~T Barr}, {and} \bibinfo{person}{Premkumar
  Devanbu}.} \bibinfo{year}{2014}\natexlab{}.
\newblock \showarticletitle{Comparing static bug finders and statistical
  prediction}. In \bibinfo{booktitle}{{\em Proceedings of the 36th
  International Conference on Software Engineering}}. ACM,
  \bibinfo{pages}{424--434}.
\newblock


\bibitem[\protect\citeauthoryear{Ray, Hellendoorn, Godhane, Tu, Bacchelli, and
  Devanbu}{Ray et~al\mbox{.}}{2016}]%
        {ray2016naturalness}
\bibfield{author}{\bibinfo{person}{Baishakhi Ray}, \bibinfo{person}{Vincent
  Hellendoorn}, \bibinfo{person}{Saheel Godhane}, \bibinfo{person}{Zhaopeng
  Tu}, \bibinfo{person}{Alberto Bacchelli}, {and} \bibinfo{person}{Premkumar
  Devanbu}.} \bibinfo{year}{2016}\natexlab{}.
\newblock \showarticletitle{On the naturalness of buggy code}. In
  \bibinfo{booktitle}{{\em Proceedings of the 38th International Conference on
  Software Engineering}}. ACM, \bibinfo{pages}{428--439}.
\newblock


\bibitem[\protect\citeauthoryear{Sculley}{Sculley}{2010}]%
        {Sculley:2010}
\bibfield{author}{\bibinfo{person}{D. Sculley}.}
  \bibinfo{year}{2010}\natexlab{}.
\newblock \showarticletitle{Web-scale K-means Clustering}. In
  \bibinfo{booktitle}{{\em Proceedings of the 19th International Conference on
  World Wide Web}} {\em (\bibinfo{series}{WWW '10})}. \bibinfo{publisher}{ACM},
  \bibinfo{address}{New York, NY, USA}, \bibinfo{pages}{1177--1178}.
\newblock
\showISBNx{978-1-60558-799-8}
\showDOI{%
\url{http://dx.doi.org/10.1145/1772690.1772862}}


\bibitem[\protect\citeauthoryear{Storn and Price}{Storn and Price}{1997}]%
        {storn1997differential}
\bibfield{author}{\bibinfo{person}{R. Storn} {and} \bibinfo{person}{K. Price}.}
  \bibinfo{year}{1997}\natexlab{}.
\newblock \showarticletitle{Differential evolution--a simple and efficient
  heuristic for global optimization over continuous spaces}.
\newblock \bibinfo{journal}{{\em Journal of global optimization\/}}
  \bibinfo{volume}{11}, \bibinfo{number}{4} (\bibinfo{year}{1997}),
  \bibinfo{pages}{341--359}.
\newblock


\bibitem[\protect\citeauthoryear{Tantithamthavorn, McIntosh, Hassan, and
  Matsumoto}{Tantithamthavorn et~al\mbox{.}}{2016}]%
        {tantithamthavorn2016automated}
\bibfield{author}{\bibinfo{person}{Chakkrit Tantithamthavorn},
  \bibinfo{person}{Shane McIntosh}, \bibinfo{person}{Ahmed~E Hassan}, {and}
  \bibinfo{person}{Kenichi Matsumoto}.} \bibinfo{year}{2016}\natexlab{}.
\newblock \showarticletitle{Automated parameter optimization of classification
  techniques for defect prediction models}. In \bibinfo{booktitle}{{\em
  Software Engineering (ICSE), 2016 IEEE/ACM 38th International Conference
  on}}. IEEE, \bibinfo{pages}{321--332}.
\newblock


\bibitem[\protect\citeauthoryear{Wang, Liu, and Tan}{Wang
  et~al\mbox{.}}{2016}]%
        {wang2016automatically}
\bibfield{author}{\bibinfo{person}{Song Wang}, \bibinfo{person}{Taiyue Liu},
  {and} \bibinfo{person}{Lin Tan}.} \bibinfo{year}{2016}\natexlab{}.
\newblock \showarticletitle{Automatically learning semantic features for defect
  prediction}. In \bibinfo{booktitle}{{\em Proceedings of the 38th
  International Conference on Software Engineering}}. ACM,
  \bibinfo{pages}{297--308}.
\newblock


\bibitem[\protect\citeauthoryear{Witten and Frank}{Witten and Frank}{2002}]%
        {Witten:2002}
\bibfield{author}{\bibinfo{person}{Ian~H. Witten} {and} \bibinfo{person}{Eibe
  Frank}.} \bibinfo{year}{2002}\natexlab{}.
\newblock \showarticletitle{Data Mining: Practical Machine Learning Tools and
  Techniques with Java Implementations}.
\newblock \bibinfo{journal}{{\em SIGMOD Rec.\/}} \bibinfo{volume}{31},
  \bibinfo{number}{1} (\bibinfo{date}{March} \bibinfo{year}{2002}),
  \bibinfo{pages}{76--77}.
\newblock
\showISSN{0163-5808}
\showDOI{%
\url{http://dx.doi.org/10.1145/507338.507355}}


\bibitem[\protect\citeauthoryear{Witten and Frank}{Witten and Frank}{2005}]%
        {Witten:2005}
\bibfield{author}{\bibinfo{person}{Ian~H. Witten} {and} \bibinfo{person}{Eibe
  Frank}.} \bibinfo{year}{2005}\natexlab{}.
\newblock \bibinfo{booktitle}{{\em Data Mining: Practical Machine Learning
  Tools and Techniques, Second Edition (Morgan Kaufmann Series in Data
  Management Systems)}}.
\newblock \bibinfo{publisher}{Morgan Kaufmann Publishers Inc.},
  \bibinfo{address}{San Francisco, CA, USA}.
\newblock
\showISBNx{0120884070}


\bibitem[\protect\citeauthoryear{Yang, Zhou, Liu, Zhao, Lu, Xu, Xu, and
  Leung}{Yang et~al\mbox{.}}{2016}]%
        {yang2016effort}
\bibfield{author}{\bibinfo{person}{Yibiao Yang}, \bibinfo{person}{Yuming Zhou},
  \bibinfo{person}{Jinping Liu}, \bibinfo{person}{Yangyang Zhao},
  \bibinfo{person}{Hongmin Lu}, \bibinfo{person}{Lei Xu},
  \bibinfo{person}{Baowen Xu}, {and} \bibinfo{person}{Hareton Leung}.}
  \bibinfo{year}{2016}\natexlab{}.
\newblock \showarticletitle{Effort-aware just-in-time defect prediction: simple
  unsupervised models could be better than supervised models}. In
  \bibinfo{booktitle}{{\em Proceedings of the 2016 24th ACM SIGSOFT
  International Symposium on Foundations of Software Engineering}}. ACM,
  \bibinfo{pages}{157--168}.
\newblock


\bibitem[\protect\citeauthoryear{Yoo and Harman}{Yoo and Harman}{2012}]%
        {yoo2012regression}
\bibfield{author}{\bibinfo{person}{Shin Yoo} {and} \bibinfo{person}{Mark
  Harman}.} \bibinfo{year}{2012}\natexlab{}.
\newblock \showarticletitle{Regression testing minimization, selection and
  prioritization: a survey}.
\newblock \bibinfo{journal}{{\em Software Testing, Verification and
  Reliability\/}} \bibinfo{volume}{22}, \bibinfo{number}{2}
  (\bibinfo{year}{2012}), \bibinfo{pages}{67--120}.
\newblock


\bibitem[\protect\citeauthoryear{Zhang, Zheng, Zou, and Hassan}{Zhang
  et~al\mbox{.}}{2016}]%
        {zhang2016cross}
\bibfield{author}{\bibinfo{person}{Feng Zhang}, \bibinfo{person}{Quan Zheng},
  \bibinfo{person}{Ying Zou}, {and} \bibinfo{person}{Ahmed~E Hassan}.}
  \bibinfo{year}{2016}\natexlab{}.
\newblock \showarticletitle{Cross-project defect prediction using a
  connectivity-based unsupervised classifier}. In \bibinfo{booktitle}{{\em
  Proceedings of the 38th International Conference on Software Engineering}}.
  ACM, \bibinfo{pages}{309--320}.
\newblock


\end{thebibliography}

\end{document}